# Joint Program Partitioning and Resource Allocation for Completion Time Minimization in Multi-MEC Systems

Taizhou Yi, Guopeng Zhang, Kezhi Wang and Kun Yang

*Abstract*—This paper considers a practical mobile edge computing (MEC) system, where edge server does not pre-install the program required to perform user offloaded computing tasks. A *partial program offloading* (PPO) scheme is proposed, which can divide a user program into two parts, where the first part is executed by the user itself and the second part is transferred to an edge server for remote execution. However, the execution of the latter part requires the results of the previous part (called *intermediate result*) as the input. We aim to minimize the overall time consumption of a multi-server MEC system to complete all user offloaded tasks. It is modeled as a mixed integer nonlinear programming (MINLP) problem which considers *user-and-server association*, *program partitioning*, and *communication resource allocation* in a joint manner. An effective algorithm is developed to solve the problem by exploiting its structural features. First, the task completion time of a single server is minimized given the computing workload and available resource. Then, the working time of the edge servers are balanced by updating user-and-server association and communication resource allocation. Numerical results show that significant performance improvement can be achieved by the proposed scheme.

## I. Introduction

The development of mobile Artificial Intelligence (AI) and Machine Learning (ML) applications lead to the advancement of the hardware for user equipment (UE) [1]. Although nowadays, some AI applications, e.g., ResNet [2] and MobileNet [3] may be handled by UEs locally, other computational-intensive applications, e.g., augmented reality (AR) [4] services may still require to be processed in the central cloud. However, cloud normally requires mobile users to upload the tasks data before they can be processed. If too many users request computing resources at the same time, serious congestion may be incurred. To address this issue, mobile edge computing (MEC) has been proposed to deploy computing resource on the networks edge, which greatly reduces the transmission delay and network congesting.

MEC normally assumes that all the users' programs and running environments should be pre-installed on edge servers in the form of virtual machines (VMs) [5] or mobile clones [6]. Recently, with the development of network function virtualization (NFV), flexible storage and service resource scheduling have made MEC more widely deployed in computation-centric networks [7] [8] [9]. However, task offloading in MEC may face the problem of *user and server randomness*. To be more specific, the *user randomness* problem refers to that an edge server does not know which users may offload which kind of computing tasks, so that this edge server cannot pre-cache the programs needed to execute that task. On the other hand, the *server randomness* problem refers to the fact that when a group of users face several temporarily deployed edge servers (e.g., unmanned aerial vehicles (UAV) or unmanned ground vehicles (UGV)-enabled MEC [10]). In this case, it is impractical to assume that the programs required to be executed is pre-installed on these temporary servers. Under the condition that a user may execute a program several times, an effective way to address the *user and server randomness* problem is to allow the user to transfer the program to an edge server before sending the task data to the server. Although the communication and storage costs caused by this approach are high, a program file needs to be uploaded only once and can be re-run several times.

For instance, we consider a man with poor eyesight wears smart glasses (without powerful GPU) and rides the subway. In the crowded subway station, he can get help from his smart glasses to identify objects around him. Fortunately, edge servers are deployed over the station (like Wi-Fi Hotspots), so the smart glasses can use them to accelerate the object recognition service by offloading complex DNN computations to a nearby edge server. The problems caused by the above instance includes: (1) The programs required to execute user tasks are not pre-installed on the edge servers. Thus the user needs to offload the program files to the edge server first; (2) It takes too much time to offload the whole program file at one time, so partial offloading strategy is much more practical. (3) The considered neural network-based programs have layered features and the layers must be performed in a linear order. Therefore, in addition to uploading task input data, a user should transfer the *intermediate results* of its local execution to an edge server.

In this paper, we aim to design a task offloading scheme to solve the *user and server randomness* problem in MEC. We consider a multi-user and multi-server environment, where we aim to reduce the overall time required for the edge servers to

This work was supported by the National Natural Science Foundation of China (Grant 61971421, Grant 62071470, Grant 62132004, Grant 61871076 and Grant 61620106011), MOST Major Research and Development Project (Grant No.: 2021YFB2900200) and EU H2020 Project COSAFE (GA-824019). (*Corresponding author: Guopeng Zhang.*)

Taizhou Yi and Guopeng Zhang are both with the School of Computer Science and Technology, China University of Mining and Technology, Xuzhou, China, 221116. (email: taizhouyi@cumt.edu.cn, gpzhang@cumt.edu.cn)

Kezhi Wang is with the Department of Computer and Information Science, Northumbria University, Newcastle NE2 1XE, UK. (email: kezhi.wang@northumbria.ac.uk)

Kun Yang is with Zhongshan College, University of Electronic Science and Technology of China (UESTC), Zhongshan, China, 528400, also with the School of Communication and Information Engineering, University of Electronic Science and Technology of China (UESTC), Chengdu, China, 611731 and with the School of Computer Science and Electronic Engineering, University of Essex, Colchester CO4 3SQ, U.K. (kunyang@essex.ac.uk)

complete all the user offloaded tasks. To make full use of the computing resource, we assume that a user program can be partitioned into two parts of different sizes. The first part is executed by the UE itself, while the second part can be offloaded to an edge server for remote execution. We refer to this type of task offloading approach as *partial program offloading* (PPO).

In the literature, PPO methods can be divided into the following two categories according to the internal execution order of a program. For some applications, such as data compression and decompression, the execution of one part of a program does not require the result of another part as the input, so that the offloaded and un-offloaded parts of the program can be executed simultaneously at a UE and an edge server. The PPO method applied to such applications is referred to as the *parallel executable* PPO (PE-PPO) [13], as it does not consider the execution order of each part of the program. For another type of applications, such as deep neural networks (DNNs), the execution of the latter part of a program requires the result of the previous part as the input. The PPO method applied to such applications is referred to as the *sequentially executed* PPO (SE-PPO) [12], as it specifies a strict execution order for the un-offloaded and offloaded parts of a program. In this paper, we consider the SE-PPO approach. The main challenges include:

⬥ The proposed scheme requires a user to transmit not only the task data but also the related program to an edge server.
⬥ The output of the un-offloaded part executed at a UE is the input required by an edge server to execute the offloaded part. Because the output data cannot be ignored, the UE needs a certain amount of communication and time resource to transfer the *intermediate result* to the associated edge server.
⬥ Existing works on SE-PPO, e.g. [11] and [12], only considered one-to-one offloading scenario, which allows only one user to request computing resources from a specified edge server. Instead, this paper considers a more general situation where multiple users can simultaneously request computing resources from multiple edge servers.

Against above background, in this paper, we aim to minimize the overall time consumption of completing all the user offloaded computing tasks. We address the following issues in a joint manner. The first is *user-and-server association* which determines which server provides computing service to which user. The second is *program partitioning*, which not only determines the proportion of the offloaded part of a program but also affects the amount of data of the *intermediate results*. The third is *communication resource allocation*, which includes *bandwidth allocation* for the edge servers sharing the same channel, and *power control* for the UEs with tight energy constraint. The main contribution of this paper is summarized as follows.

1) A linear function is presented to describe the the quantitative relationship between the size of the offloaded part of a program and the amount of data in the *intermediate result*. Based on the non-orthogonal multiple access (NOMA) technology [13], a workflow model is proposed to support SE-PPO in multiuser and multi-server MEC systems.
2) We consider the case where a user devides a program into two parts and offloads the latter part of the program to an edge server for remote execution. The task completion time minimization is modeled as a mixed integer nonlinear programming (MINLP) problem, which takes into account the issues of *user-and-server association*, *program partitioning*, and *communication resource allocation* in a joint manner.
3) An effective algorithm is developed to solve the formulated problem by exploiting its structural features. We first minimize the task completion time of a single server given the computing workload and available resource, and then balance the working time of the different servers by updating user-and-server association and communication resource allocation.

The rest of this paper is organized as follows. In Sec. II, we review the related works. In Sec. III, we present the model of the considered multi-server MEC system and the workflow of the proposed PPO scheme. In Sec. IV, we formulate the task completion time minimization for the MEC system as a MINLP problem. Then, the algorithm to solve this problem is developed in Sec. V. In Sec. VI, we provide simulation results to verify the effectiveness of the proposed PPO scheme. Finally, we summarize this paper in Sec. VII.

## II. RELATED WORK

Early studies on MEC usually assumed that the programs and running environments required to perform user tasks are pre-installed on edge servers in the form of VMs or CloneCloud [14]. Bittencourt *et al.* [15] proposed a general VM architecture for caching user programs in fog computing. The VM architecture allows low-latency access to data and high processing power. Katsalis *et al.* [5] proposed a VM scheduling strategy to dynamically provide computing resources according to the predicted workloads. Xiang *et al.* [17] balanced the user requests for computing resources among multiple edge servers, aiming at minimizing the average response time for user request.

Due to the problem of *user and server randomness*, the above task offloading schemes may not be practical to MEC systems. To address this issue, Lu *et al.* [18] studied the remote loading and redirection of user computing tasks in MEC systems. The user programs are uploaded to edge servers by tracking the historic access patterns. By applying the parsing results to different user tasks, Li *et al.* [19] proposed an effective method to parse the library of mobile operating systems, which can improve the data transmission efficiency. Jiang *et al.* [20] studied user task offloading in fog computing environment. They proposed an energy-saving method to schedule user tasks to suitable edge servers by managing the computing and communication resources of all devices.

It inevitably leads to long execution delay when uploading user program to edge server as a whole. More and more researchers turn to the approach of PPO especially when dealing with computationally intensive tasks. Ren *et al.* [21] jointly optimized the computing and communication resource allocation, with the aim of minimizing the latency for task offloading in a multi-user MEC system. Wu *et al.* [22] established a task offloading framework to optimize the overall execution delay under the premise of energy savings. Qin *et al.* [23] studied the power allocation problem in a multiuser MEC system, and proposed a utility-based method to solve this problem. Zeng *et al.* [24] studied the problem of minimizing the task offloading delay in a multiuser MEC system. Under the same energy budget, NOMA can reduce the overall delay in comparison to TDMA.

With the wide application of distributed parallel systems, the *divisible load theory* has received significant attention in the past 20 years, e.g., [25]-[28]. The problems that can be solved by the *divisible load theory* contain the following two main features. First, the entire task can be divided into multiple sub-tasks that can be performed in *parallel*. This means that the sub-tasks of the parent task can be performed independently without affecting each other. Second, there is no strict internal execution order between the sub-tasks. In this paper, we study the distributed execution of special neural network-based applications. Although a completed neural network-based computational task can be divided into multiple sub-tasks, these sub-tasks can only be performed *consecutively* (i.e. not in *parallel*), since the latter one must obtain the results of the previous sub-task as its input. This means that the sub-tasks have a strict execution order.

The above methods are only applicable to the case where each part of a program can be executed in parallel. To deal with the *sequentially executed* user programs, Kuang *et al.* [29] proposed a program partition strategy which can effectively leverage the workloads for cloud server and mobile devices to achieve low-latency, low-energy consumption, and high throughput. Jeong *et al.* [11] proposed a task offloading method for DNN-based programs. By partially offloading DNN layers, the execution delay of user offloaded tasks can be reduced. Chai *et al.* [30] designed a partial program offloading method to schedule the layers of a convolutional neural network (CNN) in MEC system. In order to reduce the execution delay, the program partitioning, the CNN layer scheduling and the resource allocation has been solved in a joint manner. Against the above background, we show our system model as follows.

TABLE I
KEY NOTATIONS

| Notation | Description |
|---|---|
| $D_n$ | Size of the program file |
| $D_n^{\text{off}}$ | Size of the partial program file to be offloaded |
| $F_n$ | Computation intensity of the program file |
| $S_n$ | Size of the intermediate result |
| $k_n$ | Multiplicative coefficient of intermediate results |
| $b_n$ | Additive coefficient of intermediate results |
| $W$ | Bandwidth of overall MEC system |
| $w_i$ | Size of the bandwidth allocated to the edge server |
| $\alpha_{n,i}$ | UE-and-server association variable |
| $\bar{p}_n$ | Maximum transmit power of UE |
| $p_n^{\text{pm}}$ | Transmitted power when offloading program files |
| $p_n^{\text{ir}}$ | Transmitted power when offloading intermediate results |
| $g_{i,n}$ | The channel power gain from UE to the edge server |
| $\eta$ | Power spectral density of background Gaussian noise |
| $r_{i,n}$ | The achievable transmission rate from UE to the edge server |
| $v_i^{\text{srv}}$ | Computation capability of an edge server |
| $v_n^{\text{lc}}$ | Computation capability of UE |
| $T_i^{\text{lc}}$ | The time spent by the last UE of all UEs in single server system |
| $T_i^{\text{srv}}$ | The execution time of the server in a single server system |
| $T_i^{\text{pm}}$ | The time UE transmits the program files |
| $T_i^{\text{ir}}$ | The time UE transmits the intermediate results |
| $\rho_n$ | Energy consumption of computation-processing unit |
| $E_n$ | Maximum computation energy budget of UE |
| $T_i^{\text{ttl}}$ | The execution time of single server system |
| $\Gamma^{\text{ttl}}$ | The overall execution time of the whole MEC system |

### III. SYSTEM MODEL

The considered MEC system consists of multiple computing servers, as shown in Fig. 1. Let $\mathcal{J} = \{1,2,\dots,I\}$ denote the set of the $I$ edge servers, which can belong to different or the same service providers with overlapping coverage areas. Let $\mathcal{N} = \{1,2,\dots,N\}$ denote the set of the $N$ UEs, each of which has a compute-intensive task to be executed. Due to the limited amount of the resource, the UEs may offload the tasks partially or fully to the edge servers, to shorten the task execution time or reduce the energy consumption.

We will address the user and server randomness problem and consider that the programs required to execute user computing tasks are not pre-installed on the edge servers. Also assume that the computing task may be executed multiple times by one user. The user could upload the related program to an edge server before sending the task data to the server for remote execution. Next, we first present the task model and then introduce the communication model for data transmission. Then, we present the workflow of the proposed PPO scheme. For the reference, the variables used in this paper are listed in Table I.

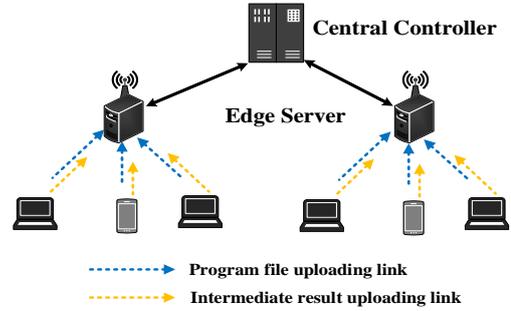

Fig. 1 Multiuser and multi-server MEC system

#### A. Computing Task Model

We consider the programs with sequential execution structure. As illustrated in Fig. 2, all the layers of the program, e.g., DNN, will be executed in a strict order, and the execution of the latter layer requires the result of the previous layer as the input. Assume the *intermediate results* produced by the previous layer is non-trivial and cannot be ignored during offloading.

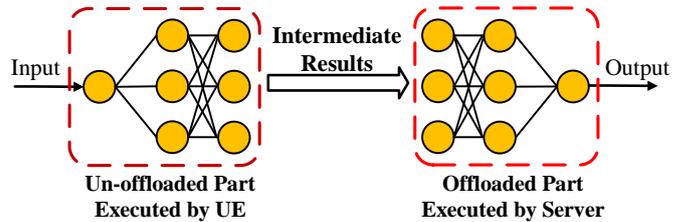

Fig. 2 The internal structure of DNN-based user programs.

We describe the program to be offloaded by the $n^{\text{th}}$ UE as a tuple $[D_n, F_n]$, where $D_n$ (in Mb) and $F_n$ (in GHz/Mb) represent the amount of data and the required computing resource of the program, respectively. The program is divided into two parts with different sizes. The first part, with size $D_n - D_n^{\text{off}}$, is executed locally at the UE, while the second part, with size $D_n^{\text{off}} \in [0, D_n]$, is offloaded to an edge server for remote execution. Since the second part of the program can be executed only when the *intermediate result* produced by the first part being obtained, we use $S_n$ to represent the amount of data of the *intermediate result*.

In this work, we focus on the application scenarios for *DNN-based* user programs, such as image/object recognition and semantic recognition. The input data of such user programs is usually a picture, a piece of voice or a paragraph of text. The input data has a fixed size and has no explicit relationship with the size of the program. In addition, according to the task offloading scheme (see Sec. III-C), the first part of a user program is performed by the related UE locally. Hence, it is not necessary to upload the task input data to an edge server. This means that the size of the task input data of a user program has no significant impact on the analysis of subsequent program offloading.

It has been shown in [30] that $S_n$ is closely related to $D_n^{\text{off}}$. We then approximate the quantitative relationship between $S_n$ and $D_n^{\text{off}}$ by the following linear function:

$$S_n = k_n D_n^{\text{off}} + b_n, \forall n \in \mathcal{N} \quad (1)$$

where $k_n$ and $b_n$ are determined by the type of the offloaded program which can be obtained by data fitting or regression methods [31]. In eq. (1) $k_n$ is a coefficient, and the unit of $b_n$ is Mb, same as $D_n^{\text{off}}$.

For example, we consider a multi-layer *back propagation neural network* (BPNN) as shown in Fig. 3(a). Let $m^{\text{in}}$ and $m^{\text{out}}$ denote the number of neurons in the input and output layers of the BPNN, respectively. According to [32], the total number of neurons in the intermediate hiding layers of the network can be obtained as $m^{\text{h}} = (m^{\text{in}} + m^{\text{out}})/2$.

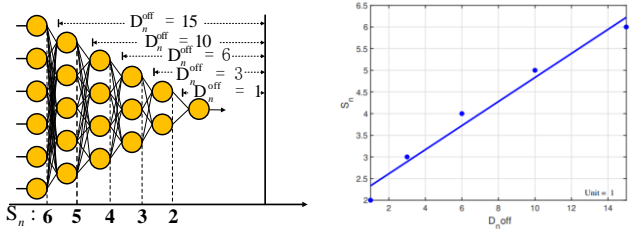

(a) Example of a multi-layer BPNN.   (b) Data fitting for $k_n$ and $b_n$.
Fig. 3 An example of a multi-layer BPNN and data fitting.

From Fig. 3(a), we can observe that the value of $S_n$ is related to the number of neurons in the kth layer, and the value of $D_n^{\text{off}}$ is related to the number of all neurons behind the kth layer. We normalize the amount of input and output data of each neuron to 1. Then we can get some specific sample points of tuple $(S_n, D_n^{\text{off}})$ as (6,15), (5,10), (4,6), (3,3) and (2,1) as shown in Fig. 3(b). By fitting these data points we can further get a line $S_n = k_n D_n^{\text{off}} + b_n$ as shown in Fig. 3(b), where $k_n = 0.2778$ and $b_n = 2.0556$.

The reason for choosing this relationship was first because BPNN is widely used and second because $D_n^{\text{off}}$ and $b_n$ in BPNN have linear relationship as shown in eq. (1), thus facilitating subsequent theoretical analysis. The unit of $b_n$ is Mb, same as $D_n^{\text{off}}$. In addition, we also note that eq. (1) is also used by other literature, e.g., [33]. The authors define a positive tuple $[I_n, C_n, O_n, T_n]$ to represent a computing task of a mobile user, where $I_n$ is the size of the task input data, and $O_n$ is the ratio of the size of the task-output data to that of the task-input data. The size of the output-data is given as $O_n I_n$. This can be seen as a special case of eq. (1).

*B. Communication Model*

We assume that the available bandwidth for the MEC system is $W$ Hz. To avoid strong mutual interference, different edge servers operate over orthogonal frequency bands. Let $w_i$ ($w_i \geq 0$, $\forall i \in \mathcal{I}$) represent the bandwidth allocated to the $i^{\text{th}}$ edge server. Then *bandwidth allocation* for the servers must meet the following constraint.

$$\sum_{i=1}^{I} w_i = W \quad (2)$$

Let $\mathbf{A} = [\alpha_{n,i}]_{N \times I}$ denote the user-and-server association matrix, where $\alpha_{n,i} \in \{0,1\}$. If the $n^{\text{th}}$ UE is associated to the $i^{\text{th}}$ edge server, then we have $\alpha_{n,i} = 1$, otherwise, $\alpha_{n,i} = 0$. Under the assumption that each UE can only choose to associate with one edge server, the user-and-server association should meet the following constraints.

$$\sum_{i=1}^{I} \alpha_{n,i} = 1, \forall n \in \mathcal{N} \quad (3)$$

The benefit of this many-to-one association method is that not only can an edge server assume the computational load of multiple UEs, but a UE needs to transmit the *intermediate results* to a certain server once. Otherwise, if a UE can offload one of its computational tasks to multiple edge servers, all the relevant servers have to work with serial mode since the different fragments of a DNN-based program are executed serial. That is, the server that carries the $k^{\text{th}}$ program fragment must wait for the server that carries the $(k-1)^{\text{th}}$ program fragment to be executed and obtain its calculation results to continue executing the $k^{\text{th}}$ program fragment that it carries. This will cause excessive wireless data transmission, thus increasing the completion delay of the computing task. This many-to-one task offloading scheme has also been used by other related works, such as [22] and [23].

We represent the set of UEs that associate with the $i^{\text{th}}$ edge server by $\mathcal{N}_i = \{n \in \mathcal{N} | \alpha_{n,i} = 1, \forall i \in \mathcal{I}\}$. Then we have $\mathcal{N}_i \subseteq \mathcal{N}$, $\cup_i \mathcal{N}_i = \mathcal{N}$, and $\mathcal{N}_i \cap \mathcal{N}_j = \emptyset$ for $\forall i,j \in \mathcal{I}$ and $i \neq j$. We assume that each edge server ranks the associated UEs in descending order according to the user-server channel gain. Then, we use $\Xi_i = (\pi_{i,1}, \ldots, \pi_{i,|\mathcal{N}_i|})$ to denote the permutation of the labels of the UEs in set $\mathcal{N}_i$, where $|\mathcal{N}_i|$ is the cardinality of $\mathcal{N}_i$. Then, for any $\pi_{i,l}{}^{\text{th}}$ ($\pi_{i,1} \leq \pi_{i,l} \leq |\mathcal{N}_i|$) UE in set $\mathcal{N}_i$, we use $p_{\pi_{i,l}}$ ($p_{\pi_{i,l}} \geq 0$) to represent the transmit power of the UE to upload data to the associated edge server.

To improve the spectrum efficiency of the system, we assume that the UEs associated with the same edge server upload their data by using the NOMA technology [34]. According to ref. [45], when successive interference cancellation (SIC) is used, the UE with good channel conditions can be decoded first to avoid the waste of energy resources. That is to decode the information of $UE_3$, $UE_2$ and $UE_1$ sequentially according to the order $g_3 < g_2 < g_1$, where $g_i$ is the channel gain from UE $i$ to the edge server. Therefore, the achievable data rate of the UE is given by

$$r_{\pi_{i,l}} = w_i \log_2 \left(1 + \frac{p_{\pi_{i,l}} g_{\pi_{i,l}}}{\sum_{j=\pi_{i,1}}^{\pi_{i,l}-1} p_{\pi_{i,j}} g_{\pi_{i,j}} + w_i N_0}\right), 1 \leq l \leq |\mathcal{N}_i| \quad (4)$$

where $g_{\pi_{i,l}}$ is the channel gain from the $\pi_{i,l}{}^{\text{th}}$ UE to the $i^{\text{th}}$ edge server, and $N_0$ is the spectral power density of the background noise.

*C. The Workflow of Program Offloading*

This paper considers a multi-user and multi-server system, and the number of users is much larger than the number of servers. If all UEs offload their computing tasks completely to edge servers, the limited computing power of edge servers may become the bottleneck of the system performance. To make full use of the computing power of the UEs and edge servers and reduce the task completion time of the system, we consider a partial offloading scheme instead of the full offloading scheme. In addition, partial offloading also improves the parallel working capability of the UEs and servers in the system, thus further reducing the time for the system to complete all the user computing tasks. In detail, we consider that each UE executes the first part of the program locally and, at the same time, uploads the second part of the program to an edge server in parallel, as shown in Fig. 4. When the edge server successfully receives the second part of the program, the intermediate result may have been prepared by the UE. The edge server can then immediately perform the second part of the program. Thus, it enhances the parallel working ability of the system and reduces the time for the system to complete all the user computing tasks.

We apply the parallel workflow in the proposed PPO scheme, as shown in Fig. 4, which includes the following four phases. We assume that the UEs associated with the same edge server are well synchronized during the operation.

Phase 1 (*Local execution and program uploading*): If the $n^{\text{th}}$ UE is associated to the $i^{\text{th}}$ edge server, it can execute the first part of the program locally while transmitting the second part to the server. Let $T_i^{\text{pm}}$ denote the time allocated to the $i^{\text{th}}$ edge server to collect the *program data* of the associated UEs. Let $T_i^{\text{lc}}$ denote the longest time to complete *local computing* among all the users associated with the $i^{\text{th}}$ edge server. Then, the time required for the $i^{\text{th}}$ edge server to complete this working phase is expressed as

$$T_i^1 = \max(T_i^{\text{lc}}, T_i^{\text{pm}}) \quad (5)$$

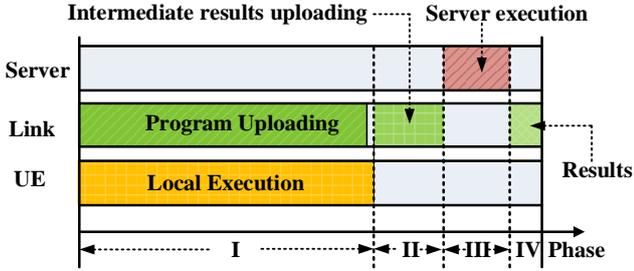

Fig. 4 The workflow of the proposed PPO scheme in a single-server scenario.

In what follows, we derive the expressions for $T_i^{\text{lc}}$ and $T_i^{\text{pm}}$.

Let $v_{\pi_{i,l}}^{\text{lc}}$ (in GHz) denote the processing speed of the computing unit of the $\pi_{i,l}^{\text{th}}$ UE. The time taken by the UE to complete computing locally is given by

$$T_{\pi_{i,l}}^{\text{lc}} = F_{\pi_{i,l}}(D_{\pi_{i,l}} - D_{\pi_{i,l}}^{\text{off}})/v_{\pi_{i,l}}^{\text{lc}}, \forall \pi_{i,l} \in \mathcal{N}_i. \quad (6)$$

By using eq. (6), we can express $T_i^{\text{lc}}$ as

$$T_i^{\text{lc}} = \max_{\pi_{i,l} \in \mathcal{N}_i} \left( \frac{F_{\pi_{i,l}}(D_{\pi_{i,l}} - D_{\pi_{i,l}}^{\text{off}})}{v_{\pi_{i,l}}^{\text{lc}}} \right), \forall i \in \mathcal{I}. \quad (7)$$

Let $p_{\pi_{i,l}}^{\text{pm}}$ denote transmit power of the $\pi_{i,l}^{\text{th}}$ UE to upload the program data. For correct and complete data reception at the $i^{\text{th}}$ edge server, we use eq. (4) to get the minimum $p_{\pi_{i,l}}^{\text{pm}}$ as [19]

$$p_{\pi_{i,l}}^{\text{pm}} = \frac{w_i N_0}{g_{\pi_{i,l}}} \delta_{\pi_{i,l}} \prod_{j=\pi_{i,1}}^{\pi_{i,l}-1} (1 + \delta_{\pi_{i,j}}), \forall \pi_{i,l} \in \mathcal{N}_i \quad (8)$$

where $\delta_{\pi_{i,l}} = \frac{p_{\pi_{i,l}} g_{\pi_{i,l}}}{\sum_{j=\pi_{i,1}}^{\pi_{i,l}-1} p_{\pi_{i,j}} g_{\pi_{i,j}} + w_i N_0}$ is the signal to interference plus noise ratio (SINR) at the receiver of the $i^{\text{th}}$ edge server. If the workload $D_n^{\text{off}}$ and the transmission time $T_i^{\text{pm}}$ are known, $\delta_{\pi_{i,l}}$ can be given by [13]

$$\delta_{\pi_{i,l}} = 2^{D_{\pi_{i,l}}^{\text{off}}/T_i^{\text{pm}} w_i} - 1, \forall \pi_{i,l} \in \mathcal{N}_i \quad (9)$$

By substituting eq. (9) into eq. (8), one can rewrite $p_{\pi_{i,l}}^{\text{pm}}$ as

$$p_{\pi_{i,l}}^{\text{pm}} = \frac{w_i N_0}{g_{\pi_{i,l}}} \left( 2^{D_{\pi_{i,l}}^{\text{off}}/T_i^{\text{pm}} w_i} - 1 \right) \times 2^{\sum_{j=\pi_{i,1}}^{\pi_{i,l}-1} D_j^{\text{off}}/T_i^{\text{pm}} w_i} \quad (10)$$

Let $\mathbf{p}_i^{\text{pm}} \triangleq \left( p_{\pi_{i,1}}^{\text{pm}}, \dots, p_{\pi_{i,|\mathcal{N}_i|}}^{\text{pm}} \right)^T \in \mathbb{R}^{|\mathcal{N}_i| \times 1}$ denote vector of the transmit powers of the UEs in set $\mathcal{N}_i$. Let $\mathbf{D}_i \triangleq$ $\left( D_{\pi_{i,1}}^{\text{off}}, \dots, D_{\pi_{i,|\mathcal{N}_i|}}^{\text{off}} \right)^T \in \mathbb{R}^{|\mathcal{N}_i| \times 1}$ denote the vector of the computing workloads uploaded by the UEs in set $\mathcal{N}_i$. We can then represent $T_i^{\text{pm}}$ as

$$T_i^{\text{pm}} = \varphi(\mathbf{p}_i^{\text{pm}}, \mathbf{D}_i) \quad (11)$$

However, the function has no explicit form. The reason why we present eq. (11) in the paper is that $T_i^{\text{pm}}$ is the variable to be optimized, and it would be better to get the analytical expressions for $T_i^{\text{pm}}$. According to eq. (10), we know that $\mathbf{p}_i^{\text{pm}}$ depends on $T_i^{\text{pm}}$ and $\mathbf{D}_i$, however, $T_i^{\text{pm}}$, $\mathbf{p}_i^{\text{pm}}$ and $\mathbf{D}_i$ are tightly coupled and the explicit form of $T_i^{\text{pm}}$ with respect to $\mathbf{p}_i^{\text{pm}}$ and $\mathbf{D}_i$ is difficult to obtain. Therefore, we give the inverse of eq. (10) as an expression for $T_i^{\text{pm}}$, namely eq. (11).

Phase 2 (*Intermediate results uploading*): This phase begins when all the UEs associated with the $i^{\text{th}}$ edge server complete their local computing and the second part of their programs is received by the server. For any $i^{\text{th}}$ edge server, we use $T_i^{\text{ir}}$ to represent the duration of this working phase, which is equal to the time spent by the server to successfully decode the *intermediate results* uploaded by the associated UEs. Next, we derive the expression for $T_i^{\text{ir}}$.

Let $p_{\pi_{i,l}}^{\text{ir}}$ represent the transmit power of the $\pi_{i,l}^{\text{th}}$ UE to upload the *intermediate result*. To ensure correct and complete reception at the $i^{\text{th}}$ edge server, we use the similar method as in eq. (10) and can get the minimum $p_{\pi_{i,l}}^{\text{ir}}$ as

$$p_{\pi_{i,l}}^{\text{ir}} = \frac{w_i N_0}{g_{\pi_{i,l}}} \left( 2^{S_{\pi_{i,l}}/T_i^{\text{ir}} w_i} - 1 \right) \times 2^{\sum_{j=\pi_{i,1}}^{\pi_{i,l}-1} S_j/T_i^{\text{ir}} w_i} \quad (12)$$

Let $\mathbf{p}_i^{\text{ir}} \triangleq \left( p_{\pi_{i,1}}^{\text{ir}}, \dots, p_{\pi_{i,|\mathcal{N}_i|}}^{\text{ir}} \right)^T \in \mathbb{R}^{|\mathcal{N}_i| \times 1}$ denote the vector of the transmit powers of the UEs in set $\mathcal{N}_i$. We can represent $T_i^{\text{ir}}$ as follows

$$T_i^{\text{ir}} = \varphi(\mathbf{p}_i^{\text{ir}}, \mathbf{D}_i). \quad (13)$$

This function also has no explicit form.

Phase 3 (*Remote execution at servers*): Once an edge server collects all the *intermediate results* of the associated UEs, it can continue to execute the second part of their programs. For the $i^{\text{th}}$ edge server, the duration of this phase is represented by $T_i^{\text{srv}}$. Let $v_i^{\text{srv}}$ (in GHz) denote the processing speed of the computing unit of the $i^{\text{th}}$ edge server. Then we have

$$T_i^{\text{srv}} = \frac{\sum_{n \in \mathcal{N}_i} F_n D_n^{\text{off}}}{v_i^{\text{srv}}}, \forall i \in \mathcal{I} \quad (14)$$

Phase 4 (*Results return*): After completing all the offloaded program parts of the associated UEs, the $i^{\text{th}}$ edge server can return the final results to them. As the size of the final results is relatively small, we simply neglect the return time in the following analysis. Although, it should be noted that in some other application scenarios, for example, video rendering and text translation, the size of the final results cannot be ignored. In such a case, if the first part of a user program is offloaded to an edge server, it is not necessary to send the final result from the server to the user in the downlink channel, thus saving the data transmission time. As this PPO approach targets different application scenarios, it is worth conducting further research.

## IV. PROBLEM FORMULATION

In real-world applications, a program is usually executed several times by one or more users over a period. Although the communication and storage cost caused by program offloading is high, a user program can be uploaded once and re-run many times. If the user requires to execute the program again, it only needs to upload the task data to the edge server, instead of the program itself [11]. To avoid the transmission latency caused by uploading the whole program at one time, a user may divide it into several parts and offload them to an edge server separately. However, in this paper, we only consider to offload the program once.

In Sec. II.C, we have presented the workflow of the proposed PPO scheme in a single-server scenario. Let $T_i^{\text{ttl}}$ denote the time taken by the $i^{\text{th}}$ edge server and the associated UEs to complete this process, then one has

$$T_i^{\text{ttl}} = \max\left(T_i^{\text{lc}}, \varphi(\mathbf{p}_i^{\text{ir}}, \mathbf{D}_i)\right) + \varphi(\mathbf{p}_i^{\text{ir}}, \mathbf{D}_i) + T_i^{\text{srv}}, \forall i \in \mathcal{I} \quad (15)$$

Next, we extend the above case to the multi-server scenario, as shown in Fig. 5.

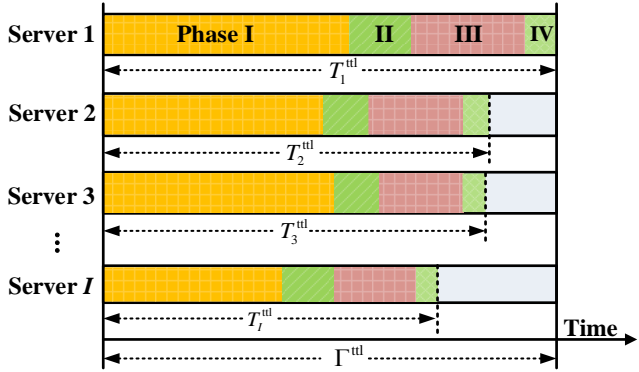

Fig. 5 The workflow of the proposed PPO scheme in a multi-server scenario.

In multi-user and multi-server MEC systems, the user-and-server association $\mathbf{A} = [\alpha_{n,i}]_{N \times I}$ determines the computation and communication workloads for each of the edge servers. The reason is:

(1) The maximum computational power $v_i^{\text{srv}}$ of any $i^{\text{th}}$ edge server can be different. In the simulation, we set $v_i^{\text{srv}}$ of each $i^{\text{th}}$ edge server between 500 GHz to 650 GHz.

(2) Even if all edge servers have the same computational capabilities, they are deployed at different places in the network. At the same time, the distribution of user devices in the network is also uneven. Different channel gains between a same UE and the edge servers will produce different communication delays, thus causing a change in the system performance. When $\mathbf{A}$ is determined, which UEs are affiliated to which edge server is known. The number of UEs associated with an edge server, the channel quality between the edge server and the associated UEs, and the amount of computational tasks offloaded by the associated UEs will have a great impact on the overall task completion time of the whole system.

Additionally, the bandwidth allocation $\mathbf{w} \triangleq (w_1, \ldots, w_I)^T \in \mathbb{R}^{I \times 1}$ determines the available communication resource for each of the servers. These factors together lead to the different times for the servers to complete the tasks.

We represent the time consumed by the system to finish all the user tasks by

$$\Gamma^{\text{ttl}} = \max_{i \in \mathcal{I}} \{T_i^{\text{ttl}}\} \quad (16)$$

which is equal to the time spent by the server which takes the longest time among all the edge servers in the system.

The system goal of the proposed PPO scheme is to minimize $\Gamma^{\text{ttl}}$, which is formulated as the following optimization problem:

$$\min_{\mathbf{D}_i, \mathbf{w}, \mathbf{A}, \{T_i^{\text{pm}}\}, \{T_i^{\text{ir}}\}} \Gamma^{\text{ttl}} \quad (17)$$

s.t. $\sum_{i=1}^{I} w_i = W$ (17.1)

$\sum_{i=1}^{I} \alpha_{n,i} = 1, \forall n \in \mathcal{N}$ (17.2)

$p_n^{\text{pm}} \leq \bar{p}_n, \forall n \in \mathcal{N}$ (17.3)

$p_n^{\text{ir}} \leq \bar{p}_n, \forall n \in \mathcal{N}$ (17.4)

$\rho_n \frac{F_n(D_n - D_n^{\text{off}})}{v_n^{\text{loc}}} \leq E_n, \forall n \in \mathcal{N}$ (17.5)

$\alpha_{n,i} \in \{0,1\}, \forall n \in \mathcal{N}, \forall i \in \mathcal{I}$ (17.6)

$0 \leq D_n^{\text{off}} \leq D_n, \forall n \in \mathcal{N}$ (17.7)

$T_i^{\text{pm}} \geq 0, \forall i \in \mathcal{I}$ (17.8)

$T_i^{\text{ir}} \geq 0, \forall i \in \mathcal{I}$ (17.9)

where constraint (17.3) indicates that the transmit power of any $n^{\text{th}}$ UE when uploading program cannot exceed the maximum power $\bar{p}_n$; Similarly, constraint (17.4) is the power constraint when any $n^{\text{th}}$ UE uploading intermediate results; $\rho_n$ denotes the power of the computation unit of the $n^{\text{th}}$ UE and due to energy limitation at UEs, constraint (17.5) indicates that the energy consumption for local computing cannot exceed the energy budget $E_n$ of the UE. Since this problem involves both continuous variables $\mathbf{D}_i$, $\mathbf{w}$, $\{T_i^{\text{pm}}\}$ and $\{T_i^{\text{ir}}\}$ and binary variables $\mathbf{A}$, it is a mixed integer nonlinear programming (MINLP) problem, which cannot be solved directly by using conventional methods.

Furthermore, it is worth noting that the storage capacity limits of edge servers are not considered here. First, the storage capacity of the currently applied lightweight edge servers is about a few hundred Gb to a few Tb [21], while the program size uploaded by UEs is mostly several Mb [12]. Therefore, the storage capacity limit of edge servers can be ignored. However, the channel capacity of the wireless communication links is very limited when compared to the size of the user program, especially when a large number of mobile users simultaneously upload their task data to the edge server. Due to the severe network congestion caused by multi-user data transmission, it inevitably leads to a large increase in the delay for completing the users' computing tasks. In other words, since the storage capacity of edge server is large enough, the server storage constraint is not considered in this paper. However, with the limited spectrum resource, communication delay is an important problem to be addressed. If we do consider the storage capacity constraint for edge servers, the following new constraint $\mathbf{D}_i^T \mathbf{A}_i \leq \Lambda_i$ (where $\Lambda_i$ is the maximum storage capacity of the $i^{\text{th}}$ server) should be added to the system problem (17). Due to the non-convexity of this constraint, it leads to a very different optimization problem with respect to problem (17). It requires new solution method and algorithm to address, and deserves further investigation in our future work.

## V. SOLUTIONS

In this section, we propose an effective algorithm to solve problem (17), as indicated below:

(1) If the bandwidth allocation $\mathbf{w}$ for the servers and the user-and-server association $\mathbf{A}$ were known, problem (17) can be decomposed into $I$ independent subproblems ($I$ is the number of the edge servers). The goal of the $i^{\text{th}}$ subproblem is to minimize the time required for the $i^{\text{th}}$ edge server to complete the computing tasks offloaded by the associated UEs. These subproblems can be solved separately without loss of any optimality.

(2) We note that the overall time consumption $\Gamma^{\text{ttl}}$ of the system depends on the longest time spent among all the edge servers. Therefore, the goal of minimizing $\Gamma^{\text{ttl}}$ is equivalent to ensuring the task completion time of each edge server as close as possible. Based on this finding, we can balance the working time of the edge servers through updating the user-and-server association $\mathbf{A}$ and the bandwidth allocation $\mathbf{w}$, and result in minimal $\Gamma^{\text{ttl}}$.

## A. Minimizing The Completion Time of A Single Server

If $\mathbf{w}$ and $\mathbf{A}$ are known, problem (17) can be converted to $I$ independent subproblems, each of which can be expressed as

$$\min_{\mathbf{D}_i, T_i^{\text{pm}}, T_i^{\text{ir}}} T_i^{\text{ttl}} \tag{18}$$

$$\text{s.t.} \quad p_{\pi_{i,l},i}^{\text{pm}} \leq \bar{p}_{\pi_{i,l}}, \forall \pi_{i,l} \in \mathcal{N}_i \tag{18.1}$$

$$p_{\pi_{i,l},i}^{\text{ir}} \leq \bar{p}_{\pi_{i,l}}, \forall \pi_{i,l} \in \mathcal{N}_i \tag{18.2}$$

$$\rho_{\pi_{i,l}} \frac{F_{\pi_{i,l}}\left(D_{\pi_{i,l}} - D_{\pi_{i,l}}^{\text{off}}\right)}{v_{\pi_{i,l}}^{\text{loc}}} \leq E_{\pi_{i,l}}, \forall \pi_{i,l} \in \mathcal{N}_i \tag{18.3}$$

$$0 \leq D_{\pi_{i,l}}^{\text{off}} \leq D_{\pi_{i,l}}, \forall n \in \mathcal{N}_i \tag{18.4}$$

$$T_i^{\text{pm}} \geq 0 \tag{18.5}$$

$$T_i^{\text{ir}} \geq 0 \tag{18.6}$$

We note that problem (18) is non-convex due to the non-convexity of constraints (18.1) and (18.2). In addition, we also note that $T_i^{\text{pm}}$ and $T_i^{\text{ir}}$ have similar mathematical forms, which means they can be solved in a similar way. Next, we decompose problem (18) into the following three subproblems. Two of them are used to find the optimal $T_i^{\text{pm}}$ and $T_i^{\text{ir}}$, and the third is to obtain the optimal $\mathbf{D}_i$.

### 1) Optimizing The Program Uploading Time $T_i^{pm}$

For any given $\mathbf{D}_i$, $T_i^{\text{pm}}$ can be minimized without compromising the optimality of the solution of problem (18). Then, one has

$$\min_{T_i^{\text{pm}}} T_i^{\text{pm}} = \varphi(\mathbf{p}_i^{\text{pm}}, \mathbf{D}_i) \tag{19}$$

$$\text{s.t.} \quad p_{\pi_{i,l},i}^{\text{pm}} \leq \bar{p}_{\pi_{i,l}}, \forall \pi_{i,l} \in \mathcal{N}_i \tag{19.1}$$

$$T_i^{\text{pm}} \geq 0 \tag{19.2}$$

Problem (19) is non-convex due to the non-convexity of constraint (19.1). To make the problem more tractable, we introduce the following parameter

$$\theta_i = \frac{1}{T_i^{\text{pm}}}, \forall i \in \mathcal{I} \tag{20}$$

By substituting eq. (20) into problem (19), we can rewrite problem (19) as

$$\max_{\theta_i} \theta_i \tag{21}$$

$$\text{s.t.} \frac{w_i N_0}{g_{\pi_{i,l}}} \left( 2^{\frac{\theta_i}{w_i} \sum_{j=\pi_{i,1}}^{\pi_{i,l}} D_j^{\text{off}}} - 2^{\frac{\theta_i}{w_i} \sum_{j=\pi_{i,1}}^{\pi_{i,l-1}} D_j^{\text{off}}} \right) \leq \bar{p}_{\pi_{i,l}}, 1 < l \leq |\mathcal{N}_i| \tag{21.1}$$

$$\theta_i \geq 0 \tag{21.2}$$

It is noted that constraint (21.1) can be separated with respect to the different UEs in set $\mathcal{N}_i$. We then define function $F_{\pi_{i,l}}(\theta_i)$ with respect to $\theta_i$ as

$$F_{\pi_{i,l}}(\theta_i) = \frac{w_i N_0}{g_{\pi_{i,l}}} \left( 2^{\frac{\theta_i}{w_i} \sum_{j=\pi_{i,1}}^{\pi_{i,l}} D_j^{\text{off}}} - 2^{\frac{\theta_i}{w_i} \sum_{j=\pi_{i,1}}^{\pi_{i,l-1}} D_j^{\text{off}}} \right) - \bar{p}_{\pi_{i,l}} \tag{22}$$

Then define

$$\theta_{\pi_{i,l}}^{\max} = \arg\min\left\{ \theta_i \geq 0 | F_{\pi_{i,l}}(\theta_i) \leq 0 \right\}, 1 < l \leq |\mathcal{N}_i| \tag{23}$$

which represents the maximum value of $\theta_i$ that satisfies constraints (21.1) and (21.2). Then, problem (21) can be transformed into

$$\theta_i^* = \min_{\pi_{i,l} \in \mathcal{N}_i} \theta_{\pi_{i,l}}^{\max}, 1 < l \leq |\mathcal{N}_i| \tag{24}$$

In order to solve problem (24), we give the following Lemma.

*Lemma 1*: $F_{\pi_{i,l}}(\theta_i)$ is a monotone increasing function and there exists one and only one root for $F_{\pi_{i,l}}(\theta_i) = 0$ within the interval $\theta_i \in (0, +\infty)$.

*Proof:* Please refer to Appendix A.

According to Lemma 1, $\theta_{\pi_{i,l}}^{\max}$ can be obtained by solving $F_{\pi_{i,l}}(\theta_i) = 0$ for $\forall n \in \mathcal{N}_i$. Since $F_{\pi_{i,l}}''(\theta_i)$ exists everywhere within the interval $\theta_i \in (0, +\infty)$, one can find the root of $F_{\pi_{i,l}}(\theta_i) = 0$ by using Newton's method. We summarize the algorithm for solving problem (19) in **Algorithm 1**, and we express the solution of problem (19) as

$$T_i^{\text{pm}} = \frac{1}{\theta_i^*} \tag{25}$$

---

**Algorithm 1** Algorithm for problem (19) and (26)

1. **Initialize:** Set $l = 1$.
2. **While** $\pi_{i,l} \leq |\mathcal{N}_i|$ **do**
3.     Use the Newton's method to find $\theta_i \in (0, +\infty)$ such that $F_{\pi_{i,l}}(\theta_i) = 0$.
4.     Update $l = l + 1$.
5. **End while**
6. **Return:** $\theta_i^* = \min_{\pi_{i,l} \in \mathcal{N}_i} \theta_{\pi_{i,l}}^{\max}$.

---

### 2) Optimizing The Intermediate Result Uploading Time $T_i^{ir}$

If $\mathbf{D}_i$ is known, one can minimize $T_i^{\text{ir}}$ without compromising the optimality of the solution of problem (18). Then, one has

$$\min_{T_i^{\text{ir}}} T_i^{\text{ir}} = \varphi(\mathbf{p}_i^{\text{ir}}, \mathbf{D}_i) \tag{26}$$

$$\text{s.t.} \quad p_{\pi_{i,l},i}^{\text{ir}} \leq \bar{p}_{\pi_{i,l}}, \forall \pi_{i,l} \in \mathcal{N}_i \tag{26.1}$$

$$T_i^{\text{tx,ir}} \geq 0 \tag{26.2}$$

Since problem (26) has the similar mathmatical form as problem (19), we use the same method to address this problem. The detailed solution is omitted here to make the paper compact.

### 3) Optimizing The User Offloading Strategy $\mathbf{D}_i$

Once $T_i^{\text{pm}}$ and $T_i^{\text{ir}}$ are known, problem (18) can be simplified to

$$\min_{\mathbf{D}_i} T_i^{\text{total}}(\mathbf{D}_i) \tag{27}$$

$$\text{s.t.} \quad p_{\pi_{i,l}}^{\text{pm}} \leq \bar{p}_{\pi_{i,l}}, \forall \pi_{i,l} \in \mathcal{N}_i \tag{27.1}$$

$$p_{\pi_{i,l}}^{\text{ir}} \leq \bar{p}_{\pi_{i,l}}, \forall \pi_{i,l} \in \mathcal{N}_i \tag{27.2}$$

$$\rho_{\pi_{i,l}} \frac{F_n(D_n - D_n^{\text{off}})}{v_n^{\text{l}}} \leq E_{\pi_{i,l}}, \forall \pi_{i,l} \in \mathcal{N}_i \tag{27.3}$$

$$0 \leq D_{\pi_{i,l}}^{\text{off}} \leq D_{\pi_{i,l}}, \forall \pi_{i,l} \in \mathcal{N}_i \tag{27.4}$$

The purpose of this problem is to optimize the program offloading strategies of the UEs associated with the same edge server. Next, we convert non-convex constraints (27.1) and (27.2) into convex ones, and then solve the problem by convex toolbox.

$$\tilde{p}_{\pi_{i,l}}^{\text{pm}}(\mathbf{D}_i;\mathbf{D}_i[t]) \triangleq \tilde{p}_{\pi_{i,l}}^{\text{pm}}(\mathbf{D}_i[t]) + \nabla \tilde{p}_{\pi_{i,l}}^{\text{pm}}(\mathbf{D}_i[t])^T(\mathbf{D}_i - \mathbf{D}_i[t]) +$$
$$\frac{1}{2}(\mathbf{D}_i - \mathbf{D}_i[t])^T \nabla^2 \tilde{p}_{\pi_{i,l}}^{\text{pm}}(\mathbf{D}_i[t])(\mathbf{D}_i - \mathbf{D}_i[t]) + \frac{\tau}{2}\|\mathbf{D}_i - \mathbf{D}_i[t]\|^2 \tag{33}$$

$$\tilde{p}_{\pi_{i,l}}^{\text{ir}}(\mathbf{D}_i;\mathbf{D}_i[t]) \triangleq \tilde{p}_{\pi_{i,l}}^{\text{ir}}(\mathbf{D}_i[t]) + \nabla \tilde{p}_{\pi_{i,l}}^{\text{ir}}(\mathbf{D}_i[t])^T(\mathbf{D}_i - \mathbf{D}_i[t]) +$$
$$\frac{1}{2}(\mathbf{D}_i - \mathbf{D}_i[t])^T \nabla^2 \tilde{p}_{\pi_{i,l}}^{\text{ir}}(\mathbf{D}_i[t])(\mathbf{D}_i - \mathbf{D}_i[t]) + \frac{\tau}{2}\|\mathbf{D}_i - \mathbf{D}_i[t]\|^2 \tag{37}$$

---

**Algorithm 2** Algorithm for problem (27)

1. **Initialize:** Set the iteration number as $t = 0$.
2. Find a feasible $\mathbf{D}_i[0]$.
3. **Repeat:**
4. For the given $\mathbf{D}_i[t]$, find the optimal $\mathbf{D}_i[t+1]$ by solving problem (38).
5. Update $\mathbf{D}_i[t+1] = \mathbf{D}_i[t] + \Delta[t]$.
6. Update $t = t + 1$.
7. **Until:** The objective value of problem (38) converges, or the maximum number of iterations ($t^{\max}$) is reached.
8. **Return:** The optimal offload strategy $\mathbf{D}_i[t]$ and the objective value for problem (27).

We first deal with constraint (27.1). Define $\sigma = \frac{w_i N_0}{g_{n,i}}$ and $\alpha = 2^{T_i^{\text{tx,pm}}/w_i}$. We can rewrite $p_{\pi_{i,l}}^{\text{pm}}$ as

$$p_{\pi_{i,l}}^{\text{pm}}(\mathbf{D}_i) = \sigma \left( \alpha^{\sum_{j=\pi_{i,1}}^{\pi_{i,l}} D_j^{\text{off}}} - \alpha^{\sum_{j=\pi_{i,1}}^{\pi_{i,l-1}} D_j^{\text{off}}} \right), \pi_{i,l} \in \mathcal{N}_i \tag{28}$$

The partial derivatives of the $p_{\pi_{i,l}}^{\text{pm}}(\mathbf{D}_i)$ with respect to $D_{\pi_{i,l}}^{\text{off}}$ ($l \in [1, |\mathcal{N}_i| - 1]$) and $D_{\pi_{i,|\mathcal{N}_i|}}^{\text{off}}$ are given by

$$\nabla_{D_{\pi_{i,l}}^{\text{off}}} p_{\pi_{i,l}}^{\text{pm}}(\mathbf{D}_i) = H_{\pi_{i,l}}\left(D_{\pi_{i,l}}^{\text{off}}\right)$$
$$= \sigma \ln \alpha \left( \alpha^{\sum_{j=\pi_{i,1}}^{\pi_{i,l}} D_{\pi_{i,l}}^{\text{off}}} - \alpha^{\sum_{j=\pi_{i,1}}^{\pi_{i,l-1}} D_{\pi_{i,l}}^{\text{off}}} \right) \tag{29}$$

and

$$\nabla_{D_{\pi_{i,|\mathcal{N}_i|}}^{\text{off}}} p_{\pi_{i,l}}^{\text{pm}}(\mathbf{D}_i) = H_{\pi_{i,|\mathcal{N}_i|}}\left(D_{\pi_{i,|\mathcal{N}_i|}}^{\text{off}}\right)$$
$$= \sigma \ln \alpha \left( \alpha^{\sum_{j=\pi_{i,1}}^{\pi_{i,l}} D_{\pi_{i,l}}^{\text{off}}} \right) \tag{30}$$

respectively. The properties of $p_{\pi_{i,l}}^{\text{pm}}(\mathbf{D}_i)$ are described by the following two Lemmas.

*Lemma 2:* $\nabla_{D_{\pi_{i,l}}^{\text{off}}} p_{\pi_{i,l}}^{\text{pm}}(\mathbf{D}_i)$ is Lipschitz continuous on $D_{\pi_{i,l}}^{\text{off}}$ ($\pi_{i,l} \in [\pi_{i,1}, |\mathcal{N}_i| - 1]$) with a constant $L_F$. Here, $L_F$ is the Lipschitz constant and the value of $L_F$ is given in the proof.

*Proof:* Please refer to Appendix B.

*Lemma 3:* $\nabla_{D_n^{\text{off}}} p_{\pi_{i,l}}^{\text{pm}}(\mathbf{D}_i)$ is Lipschitz continuous on $D_{\pi_{i,|\mathcal{N}_i|}}^{\text{off}}$ with a constant $L_C$. $L_C$ is the Lipschitz constant and its value is given in the proof.

*Proof:* Please refer to Appendix C.

According to ref. [35], if the first derivative of a non-convex function is Lipschitz continuous on the variables with the corresponding Lipschitz constants and each variable is nonempty, closed, and convex, the non-convex function can be converted to a convex one by successive convex approximation (SCA). Lemma 2 and Lemma 3 indicate that constraint (27.1) meets this rule and thus can be converted to a convex constraint by using SCA.

To apply the SCA method, we represent the number of iterations by $t$ ($t = 1, 2, \cdots$). Let $\mathbf{D}_i[t] = \{D_{\pi_{i,l}}^{\text{off}}[t], \forall \pi_{i,l} \in \mathcal{N}_i\}$ denote the offloading strategy of the UEs in the $t^{\text{th}}$ iteration. Let $\Delta[t] = \{\Delta D_{\pi_{i,l}}^{\text{off}}[t], \forall \pi_{i,l} \in \mathcal{N}_i\}$ denote the difference of the offloading strategy of the UEs between the $t^{\text{th}}$ iteration and the $(t+1)^{\text{th}}$ iteration. By using eq. (29) and (30), we can derive the gradient of $p_{\pi_{i,l}}^{\text{pm}}(\mathbf{D}_i[t])$ on $\mathbf{D}_i[t]$ as

$$\nabla p_{\pi_{i,l}}^{\text{pm}}(\mathbf{D}_i[t]) = \left( \frac{\partial p_{\pi_{i,l}}^{\text{pm}}}{\partial D_{\pi_{i,1}}^{\text{off}}[t]}, \frac{\partial p_{\pi_{i,l}}^{\text{pm}}}{\partial D_{\pi_{i,2}}^{\text{off}}[t]}, \cdots, \frac{\partial p_{\pi_{i,l}}^{\text{pm}}}{\partial D_{\pi_{i,|\mathcal{N}_i|}}^{\text{off}}[t]} \right)^T \tag{31}$$

and its Hessian Matrix as

$$\nabla^2 G_{n,i} = \begin{bmatrix} \frac{\partial^2 p_{\pi_{i,l}}^{\text{pm}}}{\partial D_1^{\text{off}}[t]^2} & \frac{\partial^2 p_{\pi_{i,l}}^{\text{pm}}}{\partial D_1^{\text{off}}[k] \partial D_2^{\text{off}}[t]} & \cdots & \frac{\partial^2 p_{\pi_{i,l}}^{\text{pm}}}{\partial D_1^{\text{off}}[k] \partial D_{N_i}^{\text{off}}[t]} \\ \frac{\partial^2 p_{\pi_{i,l}}^{\text{pm}}}{\partial D_2^{\text{off}}[t] \partial D_1^{\text{off}}[t]} & \frac{\partial^2 p_{\pi_{i,l}}^{\text{pm}}}{\partial D_2^{\text{off}}[t]^2} & \cdots & \frac{\partial^2 p_{\pi_{i,l}}^{\text{pm}}}{\partial D_2^{\text{off}}[k] \partial D_{N_i}^{\text{off}}[t]} \\ \vdots & \vdots & \ddots & \vdots \\ \frac{\partial^2 p_{\pi_{i,l}}^{\text{pm}}}{\partial D_{N_i}^{\text{off}}[t] \partial D_1^{\text{off}}[t]} & \frac{\partial^2 p_{\pi_{i,l}}^{\text{pm}}}{\partial D_{N_i}^{\text{off}}[t] \partial D_2^{\text{off}}[t]} & \cdots & \frac{\partial^2 p_{\pi_{i,l}}^{\text{pm}}}{\partial D_{N_i}^{\text{off}}[t]^2} \end{bmatrix} \tag{32}$$

According to [35], we can construct a strongly convex function to approximate $p_{\pi_{i,l}}^{\text{pm}}(\mathbf{D}_i)$ which is given in eq. (33) at the top of next page. Then the non-convex constraint (18.1) can be approximated by the following convex one

$$\tilde{p}_{\pi_{i,l}}^{\text{pm}}(\mathbf{D}_i;\mathbf{D}_i[t]) \le \bar{p}_{\pi_{i,l}} \tag{34}$$

As $\tau \ge 0$ is required in eq. (33), we set $\tau = 0$ for the sake of analysis. In addition, we set the step-size $\Delta[t]$ as

$$\Delta[t] = \frac{1}{t} \tag{35}$$

which is relevant to the number of iterations for algorithm convergence.

Now, we turn to constraint (27.2). We define $\beta = 2^{T_i^{\text{tx,ir}}/w_i}$ and represent $p_{\pi_{i,l}}^{\text{ir}}$ as $p_{\pi_{i,l}}^{\text{ir}}(\mathbf{D}_i[t]) = \sigma \left( \beta^{\sum_{i=1}^n S_n[t]} - \beta^{\sum_{i=1}^{n-1} S_n[t]} \right)$. Obviously, $p_{\pi_{i,l}}^{\text{ir}}(\mathbf{D}_i[t])$ has the same form as the $p_{\pi_{i,l}}^{\text{pm}}(\mathbf{D}_i[t])$. We can use the similar method to obtain the approximate constraint of constraint (18.2) as

$$\tilde{p}_{\pi_{i,l}}^{\text{ir}}(\mathbf{D}_i;\mathbf{D}_i[t]) \le \bar{p}_{\pi_{i,l}}, \tag{36}$$

where $\tilde{p}_{\pi_{i,l}}^{\text{ir}}$ is shown in eq. (37) at the top of this page.

Then, we have converted non-convex constraints (27.1) and (27.2) into convex ones. Also, Problem (27) is transformed into the following form

$$\min_{\mathbf{D}_i} T_i^{\text{total}}(\mathbf{D}_i) \tag{38}$$

$$\text{s.t. } \tilde{p}_{\pi_{i,l}}^{\text{pm}}(\mathbf{D}_i;\mathbf{D}_i[t]) \le \bar{p}_{\pi_{i,l}} \tag{38.1}$$

$$\tilde{p}_{\pi_{i,l}}^{\text{ir}}(\mathbf{D}_i;\mathbf{D}_i[t]) \le \bar{p}_{\pi_{i,l}} \tag{38.2}$$

$$\rho_{\pi_{i,l}} \frac{F_n(D_n - D_n^{\text{off}})}{v_n^l} \le E_{\pi_{i,l}}, \ \forall \pi_{i,l} \in \mathcal{N}_i \tag{38.3}$$

$$0 \le D_{\pi_{i,l}}^{\text{off}} \le D_{\pi_{i,l}}, \forall \pi_{i,l} \in \mathcal{N}_i \tag{38.4}$$

which is a standard convex optimization problem, and can be solved by using convex optimization tools, e.g., CVX [36].

*Lemma 4:* The optimal solution to problem (18) can be obtained by solving the three subproblems, i.e., problems (19), (26) and (27) sequentially.

*Proof:* Please refer to Appendix D.

Finaly, we summarize the algorithms for solving problems (17) and (27) in **Algorithm 2** and **Algorithm 3**, respectively.

---

**Algorithm 3** Overall algorithm to solve problem (18)

---

1. **Initialize:** Set the iteration number $\vartheta = 0$.
   Find a feasible offload strategy $\mathbf{D}^0_{n \in \mathcal{N}_i}$.
2. **Repeat:**
3.     Given offload strategy $\mathbf{D}^\vartheta_i$, we use **Algorithm 1** to compute $T^{\text{pm},\vartheta+1}_i$.
4.     Given offload strategy $\mathbf{D}^\vartheta_i$, we use **Algorithm 1** to compute $T^{\text{ir},\vartheta+1}_i$.
5.     Based on the obtained $T^{\text{pm},\vartheta+1}_i$ and $T^{\text{ir},\vartheta+1}_i$, use **Algorithm 2** to obtain $\mathbf{D}^{\vartheta+1}_i$.
6.     Update $\vartheta = \vartheta + 1$.
7. **Until**: The objective value of problem (18) converges, or the maximum number of iterations ($\vartheta^{\max}$) is reached.
8. **Return:** The optimal offload strategy $\mathbf{D}^*_i$, and the optimal duration $T^{\text{pm},*}_i$ and $T^{\text{ir},*}_i$.

---

### B. Balancing The Working Time for Different Edge Servers

In Sec. V.A, we have minimized $T^{\text{ttl}}_i$, i.e. the task completion time for each independent server. In this subsection, we further minimize $\Gamma^{\text{ttl}}$, i.e. the total task completion time of the MEC system. Since $\Gamma^{\text{ttl}}$ depends on the longest time spent among all the edge servers, the goal of minimizing $\Gamma^{\text{ttl}}$ is equivalent to making the task completion time of each server as close as possible.

It is noted that the user-and-server association $\mathbf{A}$ and the bandwidth allocation $\mathbf{w}$ jointly determine the computing and communication workloads as well as the available resources of each server [42]. Due to binary variable $\mathbf{A}$, it is difficult to obtain the optimal solution of problem (17) directly by using traditional methods. In this paper, we employ the *load balancing* method [40] [41] to design a heuristic algorithm to find the sub-optimal solution of this problem. *Load balancing* is a simple but effective method to assist multi-server systems to achieve a common goal. The approach of this paper is to minimize the overall execution time $\Gamma^{\text{ttl}}$ by keeping the execution times consistent across multiple edge servers. When the task execution time of one edge server is significantly longer than the other servers, the *excess user tasks* can be spread over the other servers to minimize $\Gamma^{\text{ttl}}$ through reallocating the communication and computing resources to the servers.

Let $\bar{i}$ represent the index of the edge server that takes the longest time to complete the offloaded computing tasks. Let $\underline{i}$ represent the index of the edge server that uses the shortest time to complete the offloaded computing tasks. The time takes by the $\bar{i}th$ server and the $\underline{i}th$ server are given by $T^{\text{ttl}}_{\bar{i}}$ and $T^{\text{ttl}}_{\underline{i}}$, respectively. Then, we can transform the objective function of problem (17) into the following form

$$\min \Phi = T^{\text{ttl}}_{\bar{i}} - T^{\text{ttl}}_{\underline{i}} \tag{39}$$

*Lemma 5:* With known UE-and-server association matrix $\mathbf{A}$, the objective function of problem (17) can be transformed into problem (39).

*Proof:* Please refer to Appendix E.

*Lemma 6:* The original problem (17) is equivalent to the $I$ independent subproblems given in problem (18) and problem (39).

*Proof:* Please refer to Appendix F.

According to the *Lemmas 4-6*, we can conclude that the proposed suboptimal algorithm can approach the optimal solution of the original problem.

Finally, we propose a heuristic algorithm to solve problem (39) which contains the following three steps.

#### 1) Initializing The Variables

The first step is to initialize variables $\mathbf{A}$ and $\mathbf{w}$. In addition to make them fall into the feasible domain, we also consider the following two factors. The *first* is to ensure a high-quality channel between the edge server and the associated UEs. We apply the *channel-gain based association* algorithm to allow the UEs to offload the programs to the edge server with the best channel condition, in order to determine the association matrix $\mathbf{A}$. The *second* is to ensure the *fairness* of bandwidth allocation among the edge servers. Therefore, we allocate the system bandwidth to the servers in proportion to the number of UEs. Then, we have

$$w_1 : w_2 : \ldots : w_I = |\mathcal{N}_1| : |\mathcal{N}_2| : \ldots : |\mathcal{N}_I|. \tag{40}$$

By applying Lemma 3 with the initialized $\mathbf{A}$ and $\mathbf{w}$, one can obtain the task completion time of any $i^{\text{th}}$ server as $T^{\text{ttl}}_i, \forall i \in \mathcal{I}$.

#### 2) User Migration

In general, the initial resource allocation $\mathbf{A}$ and $\mathbf{w}$ cannot minimize $\Phi$, i.e., the objective of problem (39). Let $\varepsilon > 0$ denote the preset threshold of $\Phi$. If $\Phi \leq \varepsilon$, it indicates that the current $\mathbf{A}$ and $\mathbf{w}$ are acceptable. The algorithm is terminated. Otherwise, if $\Phi > \varepsilon$, it indicates that the current result still has some room for optimization. In such a case, the intuition is to move some UEs associated with the $\bar{i}^{\text{th}}$ server to the $\underline{i}^{\text{th}}$ server, which can balance the time consumption of the edge servers.

To avoid instability, we move only one UE from the $\bar{i}^{\text{th}}$ server to the $\underline{i}^{\text{th}}$ server at a time. The rule for selecting the UE is

$$\pi_{\bar{i},j} = \arg\left\{g_{\pi_{\bar{i},j}} = \max\left(g_{\pi_{\bar{i},1}}, g_{\pi_{\bar{i},2}}, \ldots, g_{\pi_{\bar{i},|\mathcal{N}_{\bar{i}}|}}\right)\right\}. \tag{41}$$

According to this rule, the $j^{\text{th}}$ UE currently associated with the $\bar{i}^{\text{th}}$ server and has the best channel condition to the $\underline{i}^{\text{th}}$ server will be transferred to the $\underline{i}^{\text{th}}$ server.

After performing **Algorithm** 3 with the updated $\mathbf{A}$, the task completion time of the edge servers are updated as $\hat{T}^{\text{ttl}}_i, \forall i \in \mathcal{I}$, and the overall time consumption of the system is updated as

$$\hat{\Gamma}^{\text{ttl}} = \max(\hat{T}^{\text{ttl}}_i, \forall i \in \mathcal{I}) \tag{42}$$

Let $\hat{\Phi} = \hat{T}^{\text{ttl}}_{\bar{i}} - \hat{T}^{\text{ttl}}_{\underline{i}}$ and $\hat{\Theta} = \Gamma^{\text{ttl}} - \hat{\Gamma}^{\text{ttl}}$. For different combinations of $\hat{\Phi}$ and $\hat{\Theta}$, we can deal with the problem in the following three cases.

<u>Case 1</u>: If $\hat{\Theta} > 0$ and $\hat{\Phi} \leq \varepsilon$, it indicates that the current values of $\mathbf{A}$ and $\mathbf{w}$ meet the preset threshold. Then, the algorithm terminates and the current values of $\mathbf{A}$ and $\mathbf{w}$ are returned as the final result.

<u>Case 2</u>: If $\hat{\Theta} > 0$ and $\hat{\Phi} > \varepsilon$, it indicates that the result still has room for optimization. We continue to migrate UEs according to the rule shown in eq. (41), until the conditions in Case 1 or Case 3 are satisfied.

<u>Case 3</u>: If $\hat{\Theta} < 0$, it indicates that system performance cannot be further improved by applying the *user migration* only. The algorithm goes back to the previous step and performs the *bandwidth allocation* for the edge servers as described below.

### 3) Bandwidth Allocation for Edge Servers

Along with the *user migration* operation, one can transfer part of the bandwidth of the $\bar{i}^{\text{th}}$ to the $\underline{i}^{\text{th}}$ server, to further reduce the gap between them. The proposed *bandwidth allocation* method can be given by

$$\begin{cases} \widehat{w_{\bar{i}}} = w_{\bar{i}} + \overline{w} \\ \widehat{w_{\underline{i}}} = w_{\underline{i}} - \overline{w} \end{cases} \quad (43)$$

where $\overline{w} = w_{\underline{i}}/|N_{\underline{i}}|$ is size of the bandwidth transferred from the $\bar{i}^{\text{th}}$ to the $\underline{i}^{\text{th}}$ server, and $\widehat{w_{\bar{i}}}$ and $\widehat{w_{\underline{i}}}$ are the updated bandwidth allocation for the $\bar{i}^{\text{th}}$ and the $\underline{i}^{\text{th}}$ server, respectively.

---

**Algorithm 4** Algorithm for problem (17)

---

1. **Initialize:** Set the iteration number $\ell = 1$.
2.     Use MGO method to obtain association matrix $\mathbf{A}^0$.
3.     Set $w_1^0:\dots:w_I^0 = |\mathcal{N}_1^0|:\dots:|\mathcal{N}_I^0|$.
4.     Set maximum iteration number $\ell^{\max}$, threshold $\varepsilon$, and flag $= 0$.
5. **End Initialize**
6. **If** $\ell = 1$ **then**
7.     Use **Algorithm 3** to obtain $T_1^{\text{ttl},1},\dots,T_I^{\text{ttl},1}$.
8.     According to eq. (39) obtain $\Phi^1$.
9.     According to eq. (42) obtain $\Gamma^{\text{ttl},1}$.
10.     Update $\ell = \ell + 1$.
11.     **If** $\Phi^1 \leq \varepsilon$ **Then**
12.         End of the algorithm.
13.     **End If**
14. **Else**
15.     **While** $\ell \leq \ell^{\max}$ **Do**
16.         According to eq. (41), change UE's association strategy.
17.         Use **Algorithm 3** to obtain $T_1^{\text{ttl},\ell},\dots,T_I^{\text{ttl},\ell}$.
18.         According to eq. (39) obtain $\Phi^\ell$.
19.         According to eq. (42) obtain $\Gamma^{\text{ttl},\ell}, \Theta^\ell$.
20.         **If** $\Theta^\ell > 0$ **Then**
21.             **If** $\Phi^\ell \leq \varepsilon$ **Then**
22.                 End of the algorithm.
23.             **Else**
24.                 Update $\ell = \ell + 1$.
25.             **End If**
26.         **Else**
27.             Back up to the previous step.
28.             Set flag $= 1$.
29.         **End If**
30.         **If** flag $= 1$ **Then**
31.             According to eq. (43) allocate bandwidth.
32.             Set flag $= 0$.
33.         **End If**
34.     **End While**
35. **End If**
36. **Return:** Variables $\mathbf{D}, \mathbf{w}, \mathbf{A}, \{T_i^{\text{pm}}\}, \{T_i^{\text{ir}}\}$ and the objective value $\Gamma^{\text{ttl}}$ for problem (17).

---

After performing the bandwidth allocation, we can execute **Algorithm 3** again by using the newly obtained $\mathbf{w}$. According to the obtained $\widehat{\Theta}$, one can deal with the problem in the following two cases.

*Case 4*: if $\widehat{\Theta} > 0$, it indicates that the current bandwidth allocation is effective and one can continue the user migration operation as presented above.

*Case 5*: if $\widehat{\Theta} < 0$, it indicates that the current bandwidth allocation deteriorates the system performance. The algorithm can go back to the previous step and performs the bandwidth allocation (43) by using an updated $\overline{w}$.

In Case 5, the challenge is to get a new $\overline{w}$ in order to improve the system performance. Next, we address this issue. We first give the following lemma.

*Lemma 7:* After performing bandwidth allocation, if $\widehat{\Theta} < 0$, there must be $\widehat{T}_{\underline{i}}^{\text{ttl}} > \Gamma^{\text{ttl}}$, where $\Gamma^{\text{ttl}}$ is the task completion time of the system before performing the bandwidth allocation.

*Proof:* Please refer to Appendix G.

Lemma 7 shows that if the size of the bandwidth transferred from the $\bar{i}^{\text{th}}$ to the $\underline{i}^{\text{th}}$ server is too large, it will lead to the deterioration of the system performance, which means that we should use a smaller $\overline{w}$ to reallocate the bandwidth. Based on this consideration, we propose the following rule to update $\overline{w}$ as

$$\overline{w}^* = \frac{\overline{w}}{2}, \quad (44)$$

where $\overline{w}^*$ is the updated $\overline{w}$. Therefore, the bandwidth allocation can be repeated until the situation in Case 4, i.e. $\widehat{\Theta} > 0$ is satisfied.

According the above analysis, we note that Case 2 to 5 will eventually switch to Case 1. Hence, the termination condition of the designed heuristic algorithm is set as follows: the result satisfies the ordered condition in Case 1, or the execution times of the algorithm reach the pre-set threshold. Then the overall algorithm to solve problem (17) can be summarized in **Algorithm 4**.

The computational complexity of **Algorithm 4** is analyzed as follows. The Newton's method in **Algorithm 1** requires $O(1)$ to converge, i.e. the time of the algorithm is independent of the size of the input data. Since the number of times that the Newton's method is executed depends on $|\mathcal{N}_i|$, the computational complexity of **Algorithm 1** is $O(|\mathcal{N}_i|)$.

For **Algorithm 2**, since problem (27) is transformed to a convex optimization problem (38) and is solved by using CVX, the computational complexity is polynomial complexity $O(t|\mathcal{N}_i|^3)$, where $t$ is the number of iterations. The complexity of the **Algorithm 3** is $O(t|\mathcal{N}_i|^3 \log(1/\epsilon))$, where $\log(1/\epsilon)$ is the parametric of the inner loop of **Algorithm 3**.

It is pointed out in [44] that the number of iterations is related to the preset accuracy threshold of an algorithm. Let $\ell^{\max}$ denote the upper bound of the number of iterations. If the objective function to be optimized is convex, $\ell^{\max}$ is dependent on the global model's accuracy $\varepsilon$ and it can be given by $\ell^{\max} = \log(1/\varepsilon)$. It is noted that **Algorithm 4** is developed to solve problem (17), and the objective function of problem (17) is convex. Therefore, according to [44], we know that the upper bound on the number of iterations in **Algorithm 4** is $\ell^{\max} = \log(1/\varepsilon)$, and, thus **Algorithm 4** requires at most $O(\ell^{\max})$ to converge. As a result, the computational complexity of **Algorithm 4** is $O\left(t\ell^{\max}\log(1/\epsilon) \max_{i\in\mathcal{I}}|\mathcal{N}_i|^3\right)$.

### C. Algorithm Implementation

The flow chart of different problems

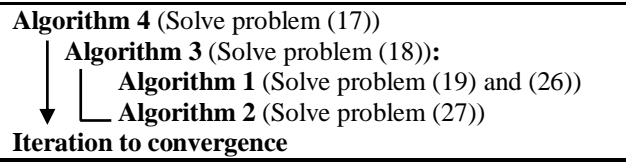

In this section, we present the details of implementing the proposed PPO scheme, i.e., **Algorithm 4**. The flow chart of the execution of **Algorithm 4** is shown above

Problem (17) is decomposed into four subproblems, that are problems (18), (19), (26), and (27). In **Algorithm 4**, **Algorithm 3** is called to solve problem (18). Then, in **Algorithm 3**, **Algorithm 1** is called to solve problem (19) and (26), and **Algorithm 2** is called to solve problem (27). Because **Algorithm 4** needs the information of system elements, such as the information of the programs to be offloaded, the computing power, the available resources of the UEs and the edge servers, as well as the channel state information of the system, we implement the PPO scheme in a centralized manner. We assume that there is a centralized controller in the MEC system that can be deployed independently or reside on an edge server. The system controller should maintain the following three tables, that are:

- The information table of the UEs, which contains parameters $v_n^{loc}$, $g_{i,n}$, $\bar{p}_n$ and $E_n$ (which $\forall n \in \mathcal{N}$).
- The information table of the edge servers, which contains $v_i^{srv}$ (which $\forall n \in \mathcal{N}$) and $\eta$.
- The information table of the programs to be offloaded, which contains $D_n$, $F_n$, $k_n$ and $b_n$ (which $\forall n \in \mathcal{N}$).

Before the users start offloading their programs, each of the UEs select the nearest edge server and request computing service from it. The edge server then forwards the user request to the system controller. After receiving the service requests of all the users, the system controller updates the *information tables* presented above and then executes **Algorithm 4**. The obtained results, including the values of $\mathbf{A}$, $\mathbf{D}_i$, $\mathbf{w}$, $\{T_i^{pm}\}$ and $\{T_i^{ir}\}$, are then sent to the edge servers and the UEs through dedicated feedback channels.

## VI. NUMERICAL RESULTS

TABLE II
Simulation Parameters

| Parameter | Value |
| --- | --- |
| Size of the application file ($D_n$) | 200 Mb |
| Computation intensity of the application file ($F_n$) | 2 GHz/M |
| Maximum transmit power of each UE ($\bar{p}_n$) | [0.1, 0.25] W |
| Maximum energy budget of each UE ($E_n$) | 4 J |
| Energy consumption of computation-processing unit ($\rho_n$) | 0.05 J |
| Computation capability of each UE ($v_n^{lc}$) | [1.2, 4] GHz |
| Computation capability of edge server ($v_i^{srv}$) | [500, 650] GHz |
| Spectral power density of background noise ($\eta$) | $10^{-20}$ W/Hz |
| Bandwidth of overall MEC system ($W$) | [4, 40] MHz |
| Coefficient of intermediate results ($k_n$) | 0.001 |
| Coefficient of intermediate results ($b_n$) | 1.5 Mb |

In this section, we show the numerical results and analyze the system performance. The simulation environment is set as a square hot-spot area of $100 \times 100$ m$^2$, where the location of the edge servers is pre-set, and location of the UEs is randomly generated. The channel gain from a UE to an edge server is set to $g_{i,n} = g_{i,0}/d_{i,n}^\kappa$, where $d_{i,n}^\kappa$ is the distance from the $n^{th}$ UE to the $i^{th}$ server, $\kappa = 3$ is the power-scaling path-loss factor, and $g_{i,0}$ follows an exponential distribution with a unit mean, which captures the fading and shadowing effects. The other simulation parameters are listed in Table II [13] [16] [30].

Since the designed **Algorithm 4** is a heuristic, in Sec. VI-A, we conduct experiment to verify whether the performance of the proposed algorithm is close to the optimal algorithm. After testifying the effectiveness of the proposed algorithm, in Sec. IV-B, we further show the performance of the proposed algorithm for large scale problems.

### A. The Effectiveness of The Proposed PPO Scheme

In this section, we show the effectiveness of the proposed PPO scheme. For that purpose, we compare the proposed algorithm with the *exhaustive search* (ES) *algorithm* and the *genetic algorithms* (GAs) in a small problem scale. Although the ES algorithm can find the optimal solution, the computational complexity is extremely high. Therefore, we assume that there are two edge servers in the system and the number of UEs is set to less than 16. Moreover, we also show the system performance when all the user programs are preinstalled on the servers. The simulation results are shown in Fig. 6.

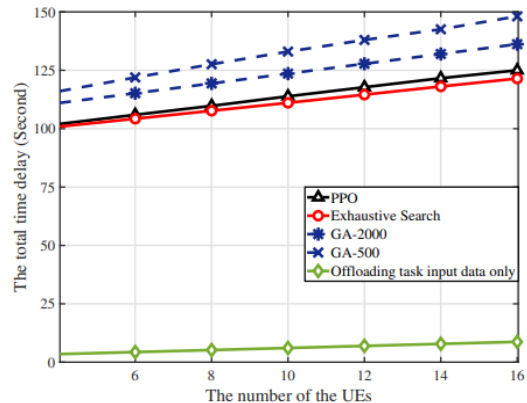

Fig. 6 The effectiveness of the proposed PPO scheme.

One can see from Fig. 6 that the performance loss of our proposed algorithm is only about 4% compared to the optimal solution obtained by using the ES algorithm, but far outperforms the performance of the GA algorithms, namely GA-500 and GA-2000 (500 and 2000 represents the number of iterations). It indicates that the proposed algorithm achieves a better compromise between computational complexity and optimal performance. In addition, one can see from Fig. 6 that a very short task completion time can be obtained when all the user programs are preinstalled on the servers. It means that after performing the proposed PPO algorithm serval rounds, the user program can be completely uploaded to the edge servers. Thereafter, the UE only needs to transfer the task input data to the server, and then achieve the performance as shown in Fig. 6 (the green curve).

Next, in Fig. 7, we show the optimization results for $D_n^{off}$ when the number of UEs is 16. Note that $D_n^{off}$ represents the data size of the offloaded part of the program of UE $n$.

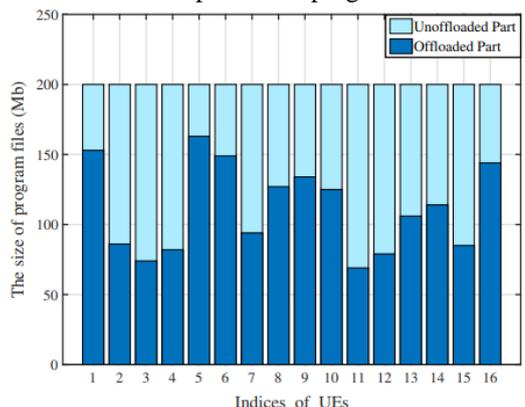

Fig. 7 The optimization results for $D_n^{off}$

From Fig. 7, one can see that although each UE is with the same amount of computing tasks, the proposed task split algorithm can optimize the offloaded parts of different UEs according to their channel conditions with the edge servers and the current computational load of the edge servers.

B. *Performance of The Proposed PPO Scheme*

In this section, we show the performance of proposed PPO scheme from two aspects: the *load balancing ability* and the *optimal resource allocation performance*. For comparison purpose, we implement the following two methods. The first is the *channel-gain based association* with *fixed bandwidth allocation* (CG-FBA), which associates a UE to the edge server with the maximum channel gain and assign each server an equal channel bandwidth for data transmission. The second is the *channel-gain based association* with *variable bandwidth allocation* (CG-VBA) method, which associates a UE to the edge server with the maximum channel gain, but allocates channel bandwidth to the edge servers in proportion to the amount of computing workloads hosted by the servers. CG-FBA and CG-VBA have no load balancing ability compared with the proposed PPO scheme, but CG-VBA has the ability of resource allocation.

We randomly place 4 edge servers and 40 UEs in the simulation area. As illustrated in Fig. 5, one can see that the distribution of the UEs around each edge server is uneven. We set $v_n^{lc} = 2\text{GHz}$, $v_i^{srv} = 600 \text{ GHz}$, $\bar{p}_n = 0.2$ W, $D_n = 200$ Mb, $E_n = 4$ J, $\rho_n = 0.05$ J and $F_n = 2\text{GHz/ Mb}$. The available bandwidth of the MEC system is set to $W = 40$ MHz, and the threshold of $\Phi$ is set to $\varepsilon = 5$s, where $\Phi$ is defined in problem (39) and $\varepsilon$ is the termination condition of the iterative algorithm to minimize $\Phi$.

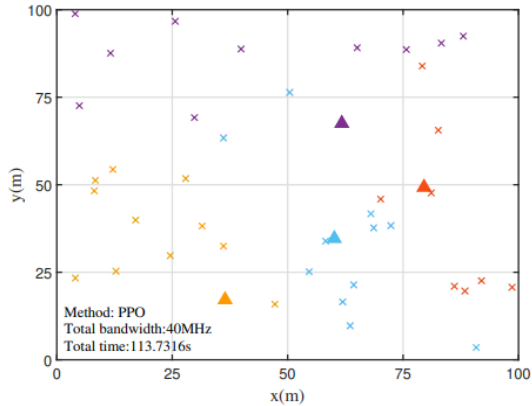

Fig. 8 User-and-server association.

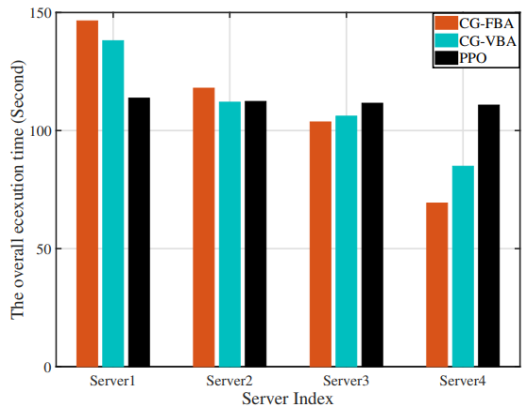

Fig. 9 Time consumption of each server.

Fig. 8 shows the user-and-server association obtained by using the proposed PPO scheme. Fig. 9 shows the time consumption of each server to complete the user offloaded tasks. In comparison to CG-FBA and CG-VBA, the proposed PPO scheme has the better performance in terms of *load balancing* as the task completion time of the servers can be well balanced. Moreover, one can see from this figure that CG-VBA reduces the difference of the time taken by the servers to complete the hosted tasks compared with CG-FBA, as CG-VAB can allocate more bandwidth to the servers with larger computing workload. However, the system goal of minimizing the overall task completion time cannot be achieved by performing *resource allocation* alone.

Next, we investigate the effect of the iterative termination parameter $\varepsilon$ on the system performance. We set $\varepsilon = 3$ and $\varepsilon = 5$ and fix $W = 20$ MHz. We set the number of the UEs as 40. One can see from Fig. 7 that when the number of edge servers increasing from 2 to 8, the execution time decreases, as expected. Then, in Fig. 8, we fix the number of the servers in the system as 4, and increase the number of UEs from 20 to 50. One can observe the similar performance as before.

In general, from Fig. 10 and Fig. 11, one can see that the proposed method outperforms CG-FBA and CG-VBA in all cases. Moreover, one can find that the proposed scheme performs better when $\varepsilon = 3$ than when $\varepsilon = 5$. This is because a smaller threshold $\varepsilon$ can reduce the time gap between any two servers in the system. However, better system performance is obtained at the expense of longer algorithm convergence time. When $\varepsilon = 5$, the execution time of the algorithm is 18.9463 seconds, while the execution time of the algorithm is 25.5821 seconds when $\varepsilon = 3$.

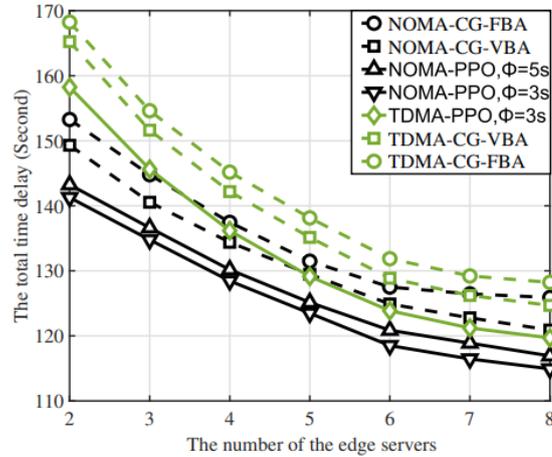

Fig. 10 The variation of system performance with the number of edge servers.

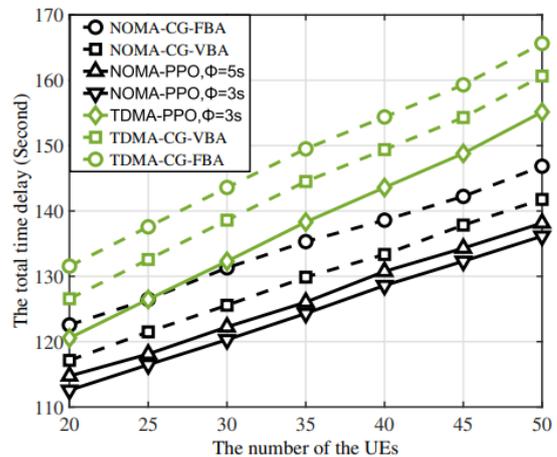

Fig. 11 The variation of system performance with the number of UEs.

Furthermore, one can see that the NOMA-based method outperforms the TDMA-based method in all cases. It indicates that although the NOMA-based method brings a higher computational complexity for transmission power control but it also obtains a higher data offloading rate than the TDMA-based method. Therefore, the time spent by the UEs to upload program data and intermediate results is greatly reduced, and the overall time consumption of the system to complete all user tasks is also greatly reduced.

*C. Effects of Available Resources on Algorithm Performance*

In this section, we study the impact of the available communication and computing resources at the UEs and edge servers. For comparisons, the following three program offloading methods are implemented. 1) *Full program offloading* (FPO): the program is offloaded as a whole to an edge server without partitioning [37]-[39]. 2) *Half program offloading* (HPO): a fixed part of 50% of a program is offloaded to an edge server. 3) *Zero program offloading* (ZPO): only local computing is conducted. Moreover, we compare the proposed algorithm with a SoA algorithm, termed as iterative heuristic MEC resource allocation (IHRA) which is proposed in [43]. The objective of IHRA is to minimize the overall task completion time, and the basic idea to assign the users in a MEC system one by one to the edge servers. In each round of the assignment, the host server allocates a certain amount of computing and communication resources to the accepted user, thus balancing the task completion time across all the servers in the system. Then, we randomly place 4 edge servers and 40 UEs in the simulated area, and set $v_n^{lc} = 2$ GHz, $v_i^{srv} = 600$ GHz, $D_n = 200$ Mb, $F_n = 2$ GHz/Mb, and $\bar{p}_n = 0.2$ W. In Fig. 12, we show the performance of the different methods when the *available system bandwidth* varies from 4 to 40 MHz.

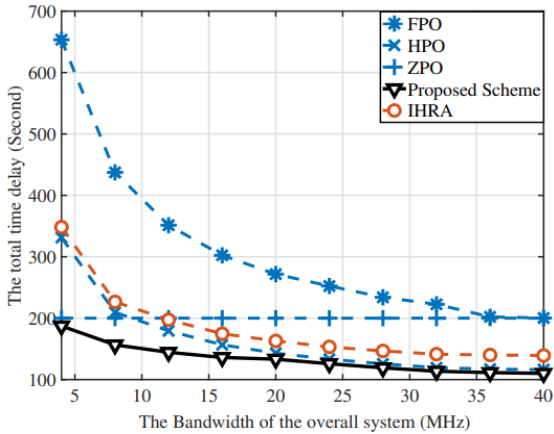

Fig. 12 The variation of system performance with the available bandwidth.

From Fig. 12, one sees that the proposed PPO scheme outperforms FPO, HPO and ZPO, as it always achieves the minimum time to complete all the tasks offloaded by the users. FPO has the worst performance because offloading a program at one time inevitably leads to a long transmission delay. When the available system bandwidth is more than 10 MHz, HPO outperforms ZPO, which implies that when the communication resource of an MEC system exceeds a certain threshold, users are encouraged to offload their programs. As the available bandwidth increases from 10 to 40 MHz, the performance of HBO improves, but still outperformed by the proposed PPO scheme. This verifies that the proposed scheme can dynamically assign user computing workloads to edge servers according to the available bandwidth resource.

Next, we check the influence of the transmit power of UE on NOMA-based communication systems. We fix the system bandwidth at $W = 20$ MHz. In Fig. 10, we show the performance of the different methods when the maximum allowable transmit power of a UE varies from 0.1 to 0.25 W.

From Fig. 13, one can see that the proposed PPO scheme always achieves the best performance. Moreover, when a low transmit power, i.e., $\bar{p}_n = 0.1$ W is applied to the UEs, the performance loss of the proposed scheme is small (about 16%) compared to using a high transmit power ($\bar{p}_n = 0.25$ W). This indicates that the proposed scheme can well control the co-channel interference caused by the transmission of the UEs connected to the same edge server.

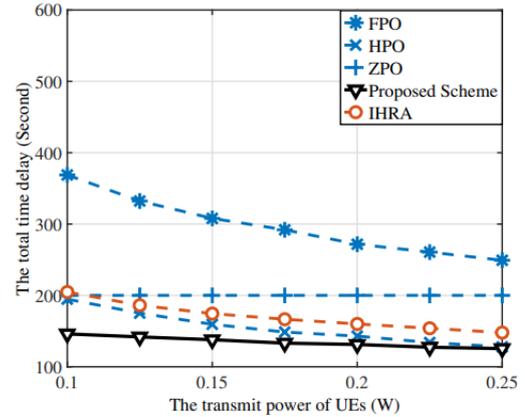

Fig. 13 The variation of system performance with the transmit power of UEs.

Next, we examine the impact of the computing power of UEs on the overall task completion time. We set $W = 20$ MHz, $\bar{p}_n = 0.2$ W, $D_n = 200$ Mb, $F_n = 2$ GHz/M, and $v_i^{srv} = 600$ GHz. In Fig. 13, we show the performance of the different methods when $v_n^{lc}$, i.e., the computing power of UEs varies from 1.2 GHz to 4 GHz.

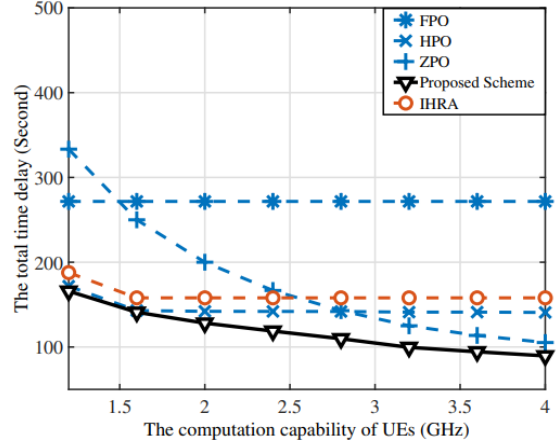

Fig. 14 The overall execution time versus computation capability of UEs.

From Fig. 14, one can see that the proposed scheme obtains the best system performance in all cases. This is because it can efficiently assign user computing workloads to the edge servers according to the computing power of the UEs. The performance of both ZPO and PPO methods gradually improve as the computing power of the UEs increases. It is because these two methods depend partly or completely on users' local computing power. In contrast, the performance of FPO is not affected by the change of UE's computing power, as it orders the users to upload the entire program to edge servers. As for HPO, the system performance cannot be improved after the computing power of the UEs exceeds a threshold of 1.6 GHz. This is because HPO always offloads a fixed proportion of user programs to edge servers. As the computing power of the edge servers remains unchanged, the bottleneck is on the server side not on the UE side.

Finally, we examine the impact of the computing power of edge servers on the system performance. We fix the computing power of each UE as 2 GHz. In Fig. 15, we show the performance of the different methods when the computing power of the servers $v_i^{\text{srv}}$ varies from 500 to 650 GHz.

From Fig. 15, one observes that the proposed method outperforms the other methods in all cases, while the change of $v_i^{\text{srv}}$ has little effect on the system performance. This is because the task execution time of the servers accounts for only a small part of the overall time consumption. It takes some time to transfer the program files and intermediate results, as expected.

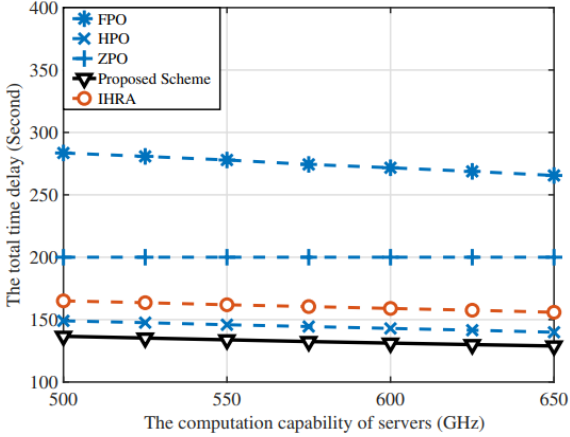

Fig. 15 The overall execution time versus computation capability of servers.

From Figs. 12-15, one can see that the proposed PPO scheme outperforms the IHRA algorithm in all cases. The main reason is that our proposed algorithm can associate multiple UEs to the corresponding edge servers from a global point of view, and, thus, make overall planning for the communication and computing resource allocation as well as the task offloading strategy of the UEs. However, the IHRA algorithm can only make system load balancing according to the resource requirement of the newly accessed UEs in each round of iteration, thus failing to obtain the global optimality.

## VII. CONCLUSION

In this paper, we have proposed a partial program offloading scheme for multi-server MEC systems. We aim to minimize the task completion time that can be modelled as a MINLP problem, which takes into account the issues of user-and-server association, program partitioning, and communication resource allocation in a joint manner. An effective algorithm has been developed to solve the problem by exploiting its structural features. The performance of the proposed scheme has been verified through extensive simulations.

## APPENDIX A

*Proof of Lemma 1*: Let $\rho = \frac{w_i N_0}{g_{\pi_{i,l}}}$, $\mu = 2^{\theta_i \sum_{j=\pi_{i,1}}^{\pi_{i,l}} D_j^{\text{off}}/w_i}$ and $\eta = 2^{\theta_i \sum_{j=\pi_{i,1}}^{\pi_{i,l}-1} D_j^{\text{off}}/w_i}$. Hence the function (22) can be expressed as $F_{\pi_{i,l}}(\theta_i) = \rho(\mu^{\theta_i} - \eta^{\theta_i}) - \bar{p}_{\pi_{i,l}}$. Its first derivative is

$$F'_{\pi_{i,l}}(\theta_i) = \rho(\mu^{\theta_i}\ln\mu - \eta^{\theta_i}\ln\eta) > 0 \quad (45)$$

therefore, the function (23) is a monotone increasing function. Because $\lim_{\theta_i \to 0} F_{\pi_{i,l}}(\theta_i) = -\bar{p}_{\pi_{i,l}} < 0$, there must be one and only one root of the function $F_{\pi_{i,l}}(\theta_i)$ within the interval $\theta_i \in (0, +\infty)$.

## APPENDIX B

*Proof of Lemma 2:* For the sake of analysis, assume $D_{\pi_{i,l}}^{\text{off}} = x$. Based on the graph of the exponential function, $H_{\pi_{i,l}}(x)$ can be seen monotonically increasing, i.e.

$$\lim_{x \to 0} H_{\pi_{i,l}}(x) \leq H_{\pi_{i,l}}(x) \leq \lim_{x \to D_j} H_{\pi_{i,l}}(x) \quad (46)$$

Hence, for any $x_1, x_2 \in [0, D_{\pi_{i,l}}]$, one can obtain

$$\left|H_{\pi_{i,l}}(x_1) - H_{\pi_{i,l}}(x_2)\right| \leq H_{\pi_{i,l}}(D_{\pi_{i,l}}) - H_{\pi_{i,l}}(0) \quad (47)$$

Furthermore, it is easy to prove that

$$|x_1 - x_2| \leq D_{\pi_{i,l}} \quad (48)$$

According to inequalities (46) and (47), we can derive that

$$\left|H_{\pi_{i,l}}(x_1) - H_{\pi_{i,l}}(x_2)\right| \leq H_{\pi_{i,l}}(D_j) - H_{\pi_{i,l}}(0)$$
$$= L_F|x_1 - x_2|$$
$$= L_F D_{\pi_{i,l}}, \quad (49)$$

where $L_F = \frac{H_{\pi_{i,l}}(D_{\pi_{i,l}}) - H_{\pi_{i,l}}(0)}{D_{\pi_{i,l}}}$ is the Lipschitz constant of $H_{\pi_{i,l}}(x)$.

## APPENDIX C

*Proof of Lemma 3:* Since **Lemma 3** is proved in the same way as **Lemma 2**, the specific proof procedure refers to **Lemma 2**.

## APPENDIX D

*Proof of Lemma 4:* By observing problem (18), we note that, for any given $D_i$, $T_i^{\text{pm}}$ and $T_i^{\text{ir}}$ can be minimized respectively without compromising the optimality of the solution. Then, we can derive the following two independent subproblems

$$\min_{T_i^{\text{pm}}} T_i^{\text{pm}} = \varphi(\mathbf{p}_i^{\text{pm}}, \mathbf{D}_i), \quad (50)$$

and

$$\min_{T_i^{\text{ir}}} T_i^{\text{ir}} = \varphi(\mathbf{p}_i^{\text{ir}}, \mathbf{D}_i) \quad (51)$$

respectively. Problems (50) and (51) are problems (19) and (26) in the revised paper, respectively.

Once $T_i^{\text{pm}}$ and $T_i^{\text{ir}}$ were obtained, problem (18) can be simplified to

$$\min_{\mathbf{D}_i} T_i^{\text{total}}(\mathbf{D}_i) \quad (52)$$

which is problem (27) in the revised paper. By solving this problem, we can get the optimal $\mathbf{D}_i$. Next, we present how to solve problem (18).

We first rewrite the expression of $T_i^{\text{ttl}}$, i.e., the objective function of problem (18) as

$$T_i^{\text{ttl}} = \max(T_i^{\text{lc}}, T_i^{\text{pm}}) + T_i^{\text{ir}} + T_i^{\text{srv}}, \forall i \in \mathcal{I}, \quad (53)$$

We note that $T_i^{\text{pm}}$, $T_i^{\text{ir}}$ and $\mathbf{D}_i$ are all convex with respect to $T_i^{\text{lc}}$, $T_i^{\text{pm}}$, $T_i^{\text{ir}}$ and $T_i^{\text{srv}}$. Since the maximum and summation operations in eq. (53) are convexity preserving operations. So the objective function of problem (18) is convex.

According to [42], since the objective function is convex, we can use the block coordinate descent (BCD) method to achieve the global optimum of problem (18). This completes the proof.

## APPENDIX E

*Proof of Lemma 5:* we first consider a simple two-server case. Then we extend it to the multi-server case.

*Two-server case*: In this case, the objective function of problem (17) can be given as

$$\min_{\mathbf{A},\mathbf{w}} \Gamma^{\text{ttl}} = \max_{i \in \mathcal{J}}\{T_1^{\text{ttl}}, T_2^{\text{ttl}}\} \quad (54)$$

By applying the *channel-gain based association* algorithm, the initial UE-and-server association matrix $\mathbf{A}$ is determined. In addition, the system bandwidth $\mathbf{w}$ is allocated to the servers in proportion to the number of accepted UEs. Thereafter, each of the servers can use **Algorithm 3** (please refer to Sec. V.A of the revised paper for details) to minimize $T_i^{\text{ttl}}$ ($i$=1, 2), i.e., the task completion time of the $i$th server.

For the sake of analysis, we assume that $T_1^{\text{ttl}} > T_2^{\text{ttl}}$ after performing **Algorithm 3**. Then, by using the *load balancing* method, we need to update $\mathbf{A}$ and $\mathbf{w}$ in the direction of minimizing $\Gamma^{\text{ttl}}$. That is to decrease $T_1^{\text{ttl}}$ and increase $T_2^{\text{ttl}}$ until $T_2^{\text{ttl}} \approx T_1^{\text{ttl}}$.

Let $\Phi_{1,2} = |T_1^{\text{ttl}} - T_2^{\text{ttl}}|$. We can transform problem (54) into the following form

$$\min_{\mathbf{A},\mathbf{w}} \Phi_{1,2} = |T_1^{\text{ttl}} - T_2^{\text{ttl}}| \quad (55)$$

This completes the proof for the two-server case.

*Multi-server case*: In such a case, we define

$$T_{\bar{i}}^{\text{ttl}} = \max_{i \in \mathcal{J}}\{T_i^{\text{ttl}}\}, \quad (56)$$

where $\bar{i}$ is the index of the edge server that takes the longest time to complete the user offloaded tasks. Then problem (17) can be rewritten as

$$\min_{\mathbf{A},\mathbf{w}} \Gamma^{\text{ttl}} = \left\{T_{\bar{i}}^{\text{ttl}}, \max_{i \in \mathcal{J}, i \neq \bar{i}}\{T_i^{\text{ttl}}\}\right\}. \quad (57)$$

By applying the *load balancing* method, problem (57) can be transformed into

$$\min_{\mathbf{A},\mathbf{w}} \Phi_{\bar{i},\{i \in \mathcal{J}, i \neq \bar{i}\}} = T_{\bar{i}}^{\text{ttl}} - T_{\{i \in \mathcal{J}, i \neq \bar{i}\}}^{\text{ttl}}, \quad (58)$$

where $T_{\{i \in \mathcal{J}, i \neq \bar{i}\}}^{\text{ttl}}$ is the task execution time for the multi-server system after eliminating the $\bar{i}$th server.

By investigating problem (58), we note that $T_{\{i \in \mathcal{J}, i \neq \bar{i}\}}^{\text{ttl}}$ can be transformed recursively by using eq. (56) and (57) until only two servers are left in the system. The multi-server problem (17) can be finally transformed into the following form

$$\min_{\mathbf{A},\mathbf{w}} \Phi_{\bar{i},\underline{i}} = T_{\bar{i}}^{\text{ttl}} - T_{\underline{i}}^{\text{ttl}}, \quad (59)$$

where $\underline{i}$ represents the index of the edge server that uses the shortest time to complete the user offloaded tasks. This completes the proof.

## APPENDIX F

*Proof of Lemma 6:* Once $\mathbf{A}$ and $\mathbf{w}$ were known, one can simplify the objective function of problem (17) as

$$\min_{\mathbf{D}_i,\{T_i^{\text{pm}}\},\{T_i^{\text{ir}}\}} \Gamma^{\text{ttl}} = \max_{i \in \mathcal{J}}\{T_i^{\text{ttl}}\} \quad (60)$$

It is obviously that the minimization of $T_i^{\text{ttl}}$ for each server $i$ through optimizing $\mathbf{D}_i$, $\{T_i^{\text{pm}}\}$ and $\{T_i^{\text{ir}}\}$ does not affect the execution time of the other servers. Therefore, the objective function of problem (17) can be decomposed into the following $I$ independent subproblems

$$\min_{\mathbf{D}_i,T_i^{\text{pm}},T_i^{\text{ir}}} T_i^{\text{ttl}} \quad (61)$$

as shown in problem (18).

Thereafter, we use problem (39) to obtain the optimal UE-and-server association matrix $\mathbf{A}$ and the optimal bandwidth allocation $\mathbf{w}$ for the $I$ servers. Then, we can conclude that the original problem (17) can be transform into problems (18) and (39) without compromising the optimality of the solution of problem (17). This completes the proof.

## APPENDIX G

*Proof of Lemma 7:* Obviously, according to eq. (4) and eq. (13), $T_i^{\text{ttl}}$ decreases as $w_i$ increases. Due to eq. (43), one can obtain

$$\begin{cases} \hat{T}_{\bar{i}}^{\text{ttl}} < T_{\bar{i}}^{\text{ttl}} \\ \hat{T}_{\underline{i}}^{\text{ttl}} > T_{\underline{i}}^{\text{ttl}} \end{cases} \quad (62)$$

Then the execution time for the other servers remain the same. Hence, we prove **Lemma 7** by contradiction.

We assume that, if $\hat{\Theta} < 0$, there should be $\hat{T}_{\bar{i}}^{\text{ttl}} < \Gamma^{\text{ttl}}$ after reallocating bandwidth. Because the execution time for the other servers remain the same, there should be

$$\hat{T}_{\underline{i}}^{\text{ttl}} = \hat{\Gamma}^{\text{ttl}} > \Gamma^{\text{ttl}}. \quad (63)$$

According to eq. (44), one has

$$T_{\bar{i}}^{\text{ttl}} = \Gamma^{\text{ttl}}. \quad (64)$$

Then we can derive that

$$\hat{T}_{\underline{i}}^{\text{ttl}} > T_{\bar{i}}^{\text{ttl}}. \quad (65)$$

This is in contradiction with eq. (62). Thus, the null hypothesis is not valid and **Lemma 7** is valid.

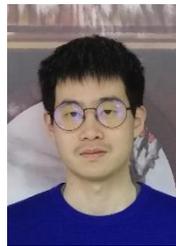

**Taizhou Yi** received the bachelor's degree from the School of Information Engineering, Xiangtan University, China, in 2019. He is currently pursuing the M.E. degree with the School of Computer Science and Technology in China University of Mining and Technology. His research interests include mobile edge computing, deep neural network and optimization theory.

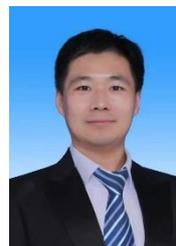

**Guopeng Zhang** received the bachelor's degree from the School of Computer Science, Jiangsu Normal University, China, in 2001, the master's degree from the School of Computer Science, South China Normal University, China, in 2005, and the Ph.D. degree from the School of Communication Engineering, Xidian University, China, in 2009. He was with ZTE Corporation Nanjing Branch for one year. In 2009, he joined the China University of Minin and Technology, China, where he is currently a Professor with the School of Computer Science and Technology. He manages research projects funded by various sources, such as the National Natural Science Foundation of China. He has published more than 60 journal and conference papers. His main research interests include wireless sensor networks, wireless personal area networks, and their applications in the Internet of Things.

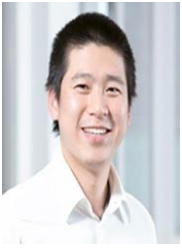

**Kezhi Wang** received his B.E. and M.E. degrees in School of Automation from Chongqing University, China, in 2008 and 2011, respectively. He received his Ph.D. degree in Engineering from the University of Warwick, U.K. in 2015. He was a Senior Research Officer in University of Essex, U.K. Currently he is a Senior Lecturer with Department of Computer and Information Sciences at Northumbria University, U.K. His research interests include wireless communications, mobile edge computing and machine learning.

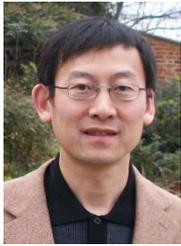

**Kun Yang** received his PhD from the Department of Electronic & Electrical Engineering of University College London (UCL), UK. He is currently a Chair Professor in the School of Computer Science & Electronic Engineering, University of Essex, leading the Network Convergence Laboratory (NCL), UK. He is also an affiliated professor at UESTC, China. Before joining in the University of Essex at 2003, he worked at UCL on several European Union (EU) research projects for several years. His main research interests include wireless networks and communications, IoT networking, data and energy integrated networks and mobile computing. He manages research projects funded by various sources such as UK EPSRC, EU FP7/H2020 and industries. He has published 150+ journal papers and filed 20 patents. He serves on the editorial boards of both IEEE (e.g., IEEE TNSE, WCL, ComMag) and non-IEEE journals. He is an IEEE ComSoC Distinguished Lecturer (2020-2021) and a Member of Academia Europaea (MAE).